\documentclass[a4paper,11pt]{article}
\pdfoutput=1

\usepackage[dvipsnames]{xcolor}
\usepackage{jheppub}

\usepackage{tikz}
\usetikzlibrary{intersections}
\usetikzlibrary{arrows,decorations.pathmorphing,backgrounds,positioning,fit,petri}
\usetikzlibrary{decorations.markings}
\usetikzlibrary{calc}
\usetikzlibrary{intersections,through,hobby}

\def\MS{\ensuremath{\overline{\mathrm{MS}}}}
\def\Lag{\ensuremath{\mathcal{L}}}
\def\Tr{\ensuremath{\mathrm{Tr}}}
\def\hc{\ensuremath{\mathrm{h.c.}}}
\def\KRp{\ensuremath{\mathrm{KR}'}}
\def\ep{\ensuremath{\varepsilon}}

\makeatletter
\newcommand{\TrR}[1]{%
T^\bgroup#1\gobblenextargT}
\newcommand{\gobblenextargT}[1]{ \textcolor{blue}{#1}\@ifnextchar\bgroup{\gobblenextarg}{\egroup }}
\newcommand{\gobblenextarg}[1]{ #1\@ifnextchar\bgroup{\gobblenextargT}{ \egroup }}

\newcommand{\TrC}[1]{%
T^\bgroup#1\gobblenextargCT}
\newcommand{\gobblenextargCT}[1]{ \textcolor{red}{#1}\@ifnextchar\bgroup{\gobblenextargC}{\egroup }}
\newcommand{\gobblenextargC}[1]{ #1\@ifnextchar\bgroup{\gobblenextargCT}{ \egroup }}

\makeatother

\title{Six-loop beta functions in general scalar theory}

\author[a,b,1]{A. Bednyakov\note{Corresponding author.}}
\author[a]{and A. Pikelner}

\affiliation[a]{
	Joint Institute for Nuclear Research, \\ 
	Joliot-Curie, 6, Dubna  141980, Russia}
\affiliation[b]{P.N. Lebedev Physical Institute of the Russian Academy of Sciences, \\
	  Leninskii pr., 5, Moscow 119991, Russia}

\emailAdd{bednya@theor.jinr.ru}
\emailAdd{pikelner@theor.jinr.ru}

\abstract{
  We consider general renormalizable scalar field theory and derive six-loop
  beta functions for all parameters in $d=4$ dimensions within the $\MS$-scheme. We do
  not explicitly compute relevant loop integrals but utilize $O(n)$-symmetric model
  counter-terms available in the literature. We consider dimensionless couplings
  and parameters with a mass scale, ranging from the trilinear self-coupling to
  the vacuum energy. We use obtained results to extend renormalization-group
  equations for several vector, matrix, and tensor models to the six-loop order. 
  Also, we apply our
  general expressions to derive new contributions to beta functions and anomalous
  dimensions in the scalar sector of the Two-Higgs-Doublet Model.
}

\begin{document} 
\maketitle
\flushbottom

\section{Introduction}
\label{sec:intro}

The renormalization group (RG) plays an essential role in high-energy physics
and the theory of critical phenomena. In particle physics, one can use RG to
re-sum specific radiative corrections making theory predictions valid in a wide
range of energy scales. In the study of critical phenomena, the RG approach
allows one to study phase transitions and predict critical exponents of the
second-order transitions with high accuracy.

A convenient tool to compute the RG functions that drive the dependence of model
parameters on the scale is to use a perturbative expansion of dimensional
regularized theory~\cite{tHooft:1973mfk} together with modified minimal \MS~subtraction of
infinities. The latter appear in loop integrals and manifest themselves in $d$
dimensions as poles in $\ep = (4-d)/2$. One cancels the poles by a finite set of
renormalization constants.

There is significant progress in the calculation of beta functions and anomalous
dimensions in the \MS~scheme. At the two-loop level, the RG functions are known
in any general renormalizable quantum field theory (QFT) in $d = 4$
dimensions~\cite{Machacek:1983fi,Machacek:1983tz,Machacek:1984zw,Luo:2002ey,Schienbein:2018fsw}.
Despite several calculations of three-loop (and even four-loop) RG functions in
particular particle-physics
models~\cite{Mihaila:2012fm,Bednyakov:2012rb,Bednyakov:2012en,Chetyrkin:2013wya,Bednyakov:2013eba,Herren:2017uxn,Bednyakov:2018cmx,Bednyakov:2015ooa,Zoller:2015tha,Chetyrkin:2016ruf,Davies:2019onf},
general three-loop results are not yet available. Recently, an essential step
has been made in this
direction~\cite{Poole:2019kcm,Steudtner:2020tzo,Steudtner:2021fzs,Thomsen:2021ncy}.
The main idea is to enumerate all possible ``tensor'' structures that can appear
in the RG functions at a certain loop level and compute the corresponding
unknown coefficients by matching them to specific models. Not long ago, this
approach allowed authors of the paper~\cite{Steudtner:2020tzo} to derive general
three-loop RG functions in a pure scalar model.

Our paper does not follow this strategy and extends the results for general
scalar theories up to six loops by more conventional technique, i.e., by
computing contributions from individual Feynman graphs. Such a leap in the loop
level is due to the significant progress in calculating critical exponents in
scalar theories. Thanks to the authors
of ref.~\cite{Kompaniets:2017yct}, the required renormalization constants can be
found given the diagram-by-diagram results of the $\KRp$ operation.
Application of the latter to a Feynman graph produces the corresponding
counter-term in the \MS~scheme.

We consider the following general renormalizable Lagrangian 
\begin{align}
	\Lag & = \frac{1}{2} \partial_\mu \phi_a \partial_\mu \phi_a - \frac{m^2_{ab}}{2} \phi_a \phi_b - \frac{h_{abc}}{3!} \phi_a \phi_b \phi_c - \frac{\lambda_{abcd}}{4!} \phi_a \phi_b \phi_c \phi_d  - t_a \phi_a - \Lambda 
	\label{eq:Lag_general}
\end{align}
for real scalar fields $\phi_a$. The mass parameters $m_{ab}^2$, cubic $h_{abc}$ and quartic couplings $\lambda_{abcd}$ are symmetric in their indices. For completeness we also add the tadpole term proportional to $t_a$, and the vacuum energy term $\Lambda$. 

Here we present the six-loop RG equations in the MS-scheme for the field 
$\phi_a$ and all parameters of eq.~\eqref{eq:Lag_general}.
The RG function for a parameter $A = \{ \lambda_{abcd}, h_{abc}, m^2_{ab}, t_{a}, \Lambda\}$ is defined as
\begin{align}
	\beta_A & \equiv  \mu\frac{ \partial A}{\partial \mu} = 
	\sum_{l} h^l \beta_A^{(l)},
	 \qquad h = \frac{1}{16 \pi^2},
\end{align}
	where $\beta_A^{(l)}$ corresponds to the $l$-loop contribution. 
	The field anomalous dimension is given by
\begin{align}
	\phi_{a,0} & = Z_{ab} \phi_b \Rightarrow \gamma^\phi_{ab} = Z_{ac}^{-1}\cdot \mu \frac{\partial Z_{cb}}{\partial \mu} 
	= - \mu \frac{\partial Z^{-1}_{ac}}{\partial \mu} \cdot Z_{cb}
	\label{eq:phi_anom_dim}
\end{align}
and is related to the field renormalization constant $Z_{ab}$. 
It is worth pointing that the latter can be multiplied by an arbitrary orthogonal matrix without spoiling divergence cancellation in two-point functions.  Due to this, the antisymmetric part of $\gamma^\phi_{ab}$ is not fixed and leads to ambiguities\footnote{We are grateful to F. Herren for bringing our attention to this fact.} in the RG functions. 
Nevertheless, the ambiguity is related to the freedom in the basis choice and does not affect physical observables (see discussions in  refs.~\cite{Bednyakov:2014pia,Herren:2017uxn}). In what follows we use symmetric $\gamma^\phi_{ab}$.

The paper is organized as follows. Section~\ref{sec:details} contains details of
our calculation. In section~\ref{sec:examples} we apply our general results
to the cases known in the literature. In particular, we consider
vector (section~\ref{sec:On}), matrix (section~\ref{sec:matrix}), and tensor
(section~\ref{sec:tensor}) models possessing different kinds of symmetries.
Also, we extend known three-loop results for the Two-Higgs-Doublet Model (2HDM)
to six loops in section~\ref{sec:thdm}. Section~\ref{sec:conclusion} contains a
discussion of the results and conclusions. In
appendix~\ref{sec:dummy_field_method} we provide a derivation of the RG
functions for dimensionful couplings in a general form.

\section{Details of calculation\label{sec:details}}

As the calculation method, we decided to use an approach similar to the one
in ref.~\cite{Kompaniets:2017yct}, based on the direct computation of the necessary
counter-terms from individual diagrams. However, in our work, we avoid the
calculation of any loop integrals. The authors of
ref.~\cite{Kompaniets:2017yct} considered all the required six-loop graphs in
the context of the $O(n)$-symmetric model\footnote{Seven-loop RG functions for the $O(n)$ model are also known due to O.~Schnetz~\cite{Schnetz:2016fhy}.} and made the corresponding counter
terms available in a computer-readable form. One can adopt the latter for more
complicated theories by changing model-dependent prefactors. In this way,
six-loop renormalization-group functions for $O(n)$ theory with cubic
anisotropy~\cite{Adzhemyan:2019gvv} and $O(n) \times O(m)$ symmetric
model~\cite{Kompaniets:2019xez} were derived.

To perform calculations with general Lagrangian (\ref{eq:Lag_general}), we
prepare a \texttt{DIANA}~\cite{Tentyukov:1999is} model file. We use special
mapping rules between its internal topology format and diagram topologies, which
are identified in ref.~\cite{Kompaniets:2017yct} and given in the Nickel index
notation. After generating all needed two- and four-point functions with
\texttt{DIANA} and performing all needed index contractions with
\texttt{FORM}~\cite{Vermaseren:1992vn}, we substitute actual values for momentum
integrals by counter-terms from the available tables~\cite{Kompaniets:2017yct}.
It is trivial to extract the RG functions $\gamma^\phi_{ab}$ and $\beta_{abcd}\equiv
\beta_{\lambda_{abcd}}$ from the first $\ep$ pole in the sum of counter-terms.

The obtained results involve a certain number of tensor structures, i.e.,
products of (up to 12) general couplings $\lambda_{abcd}$ with all but four
(two) indices contracted in $\beta_{abcd}$ ($\gamma^\phi_{ab}$). We can simplify
corresponding expressions by identifying tensor structures identical up to
the renaming of contracted indices. Also, since the corresponding numeric
coefficient depends only on the Feynman graph, we collect all the structures,
which are different only by permutations of external indices $abcd$. As a
consequence, we can cast our main result for $\beta_{abcd}$ into the form
\begin{equation}
  \label{eq:bCsTs}
  \beta_{abcd} = 
  \sum\limits_{l=1}^6 h^l \sum\limits_{i}^{n_l} T_{i,abcd}^{(l)} C_{i}^{(l)},
\end{equation}
where $n_l = \{1,2,7,23,110,571\}$ is the number of unique tensor structures
$T_{i,abcd}^{(l)}$ at $l$ loops. The coefficients $C_{i}^{(l)}$ are pure
numbers. To deduce the expressions for $T_{i,abcd}^{(l)}$, we made use of Nickel
index notation~\cite{nickel1977compilation} for graph representation of tensor
contractions and utilized the \texttt{GraphState}
package~\cite{Batkovich:2014bla}. As an example, we give here one of the
three-loop structures
\tikzset{
  internal/.style={line width=1pt,draw=black},
  dummy/.style={densely dotted,line width=1pt,draw=black}
}
\begin{equation}
  \label{eq:ts2graph}
  T^{(3)}_{4,abcd} \equiv \frac{1}{4!}\left[\lambda_{a i_1 i_2 i_3}  \lambda_{b i_3 i_4 i_6}   \lambda_{c i_2 i_4 i_5}   \lambda_{d i_1 i_5 i_6}
	  + \mathrm{perm.}
  \right]
  =
  \vcenter{\hbox{
      \begin{tikzpicture}[use Hobby shortcut, scale=0.5]
        \coordinate (vea) at (-1,1);
        \coordinate (veb) at (1,1);
        \coordinate (vec) at (1,-1);
        \coordinate (ved) at (-1,-1);
        \coordinate (via) at (-0.7,0.7);
        \coordinate (vib) at (0.7,0.7);
        \coordinate (vic) at (0.7,-0.7);
        \coordinate (vid) at (-0.7,-0.7);
        \draw[internal] (vea) -- (via);
        \draw[internal] (veb) -- (vib);
        \draw[internal] (vec) -- (vic);
        \draw[internal] (ved) -- (vid);
        \draw[internal] (via) -- (vib);
        \draw[internal] (vib) -- (vic);
        \draw[internal] (vic) -- (vid);
        \draw[internal] (vid) -- (via);
        \draw[internal] (via) -- (vic);
        \draw[internal] (vid) -- (225:0.25);
        \draw[internal] (vib) -- (45:0.25);
      \end{tikzpicture}
    }},
  \quad \mathrm{e123|e23|e3|e|}
\end{equation}
where we indicate the corresponding Nickel index and emphasize the normalization of $T^{(l)}_{i,abcd}$ together with 
the fact that the latter are symmetric in $abcd$.

We provide a table containing a minimal set of unique tensor structures formed
by different contractions between $\lambda_{abcd}$ indices and the corresponding
coefficients\footnote{Examination of the structures contributing to $\gamma^{\phi}_{ab}$ leads us to the conclusion that in pure scalar theories the ambiguity in RG functions can manifests itself starting from five loops.}. Given these tables, we derive the beta functions for dimensionful
parameters entering~\eqref{eq:Lag_general} employing the so-called dummy field
method~\cite{Martin:1993zk,Luo:2002ti,Schienbein:2018fsw}. The core of the
technique is to introduce ``dummy'' non-propagating field(s) $x_a$, e.g., by
shifting all (or just one) components of the vector $\phi_a \to \phi_a + x_a$.
Contracting $\beta_{abcd}$ with one or more dummy fields $x_a$, we can readily
obtain the expressions for $\beta_{\Lambda}$, $\beta_{a}\equiv\beta_{t_a}$,
$\beta_{ab}\equiv \beta_{m^2_{ab}}$, and $\beta_{abc}\equiv \beta_{h_{abc}}$
(see appendix~\ref{sec:dummy_field_method}). Indeed, we consider\footnote{We use
  compact notation $\beta_{xxxx} \equiv \beta_{abcd} x_a x_b x_c x_d$, etc.} $\beta_{xxxx}$,
$\beta_{axxx}$, $\beta_{abxx}$, together with $\beta_{abcx}$, and identify $\lambda_{abcx} \equiv h_{abc}$, $\lambda_{abxx} \equiv 2 m^2_{ab}$, $\lambda_{axxx} \equiv 3! t_{a}$, $\lambda_{xxxx} \equiv 4! \Lambda$.
The only subtlety here is that we have to remove contributions from external leg
renormalization, leading to tadpole diagrams in the final answer (see ref.~\cite{Schienbein:2018fsw}
for details). We can immediately identify corresponding tensor structures in
general expression for $\beta_{abcd}$

\tikzset{
  internal/.style={line width=1pt,draw=black},
  dummy/.style={densely dotted,line width=1pt,draw=black}
}

\begin{equation}
  \label{eq:se-dummy-zero}
  \vcenter{\hbox{
      \begin{tikzpicture}[use Hobby shortcut, scale=0.5]
        \coordinate (vl) at (0,0);
        \coordinate (v0) at (1.4,0);
        \coordinate (vr) at (2.8,0);
        \coordinate (v1) at (135:1);
        \coordinate (v2) at (180:0.8);
        \coordinate (v3) at (225:1);
        \coordinate (vcl) at (0.7,0);
        \coordinate (vcr) at (2.1,0);
        \draw[internal] (vl) -- (vcl);
        \draw[internal,fill=gray] (v0) circle (0.7);
        \draw[dummy] (vcr) -- (vr);
        \draw[internal] (v1) -- (vl);
        \draw[internal] (v2) -- (vl);
        \draw[internal] (v3) -- (vl);
        \fill (vl) circle (0.1);
        \fill (vcl) circle (0.1);
        \fill (vcr) circle (0.1);
      \end{tikzpicture}
    }}
  = 
  \vcenter{\hbox{
      \begin{tikzpicture}[use Hobby shortcut, scale=0.5]
        \coordinate (vl) at (0,0);
        \coordinate (v0) at (1.4,0);
        \coordinate (vr) at (2.8,0);
        \coordinate (v1) at (135:1);
        \coordinate (v2) at (180:0.8);
        \coordinate (v3) at (225:1);
        \coordinate (vcl) at (0.7,0);
        \coordinate (vcr) at (2.1,0);
        \draw[internal] (vl) -- (vcl);
        \draw[internal,fill=gray] (v0) circle (0.7);
        \draw[dummy] (vcr) -- (vr);
        \draw[internal] (v1) -- (vl);
        \draw[internal] (v2) -- (vl);
        \draw[dummy] (v3) -- (vl);
        \fill (vl) circle (0.1);
        \fill (vcl) circle (0.1);
        \fill (vcr) circle (0.1);
      \end{tikzpicture}
    }}
  =
  \vcenter{\hbox{
      \begin{tikzpicture}[use Hobby shortcut, scale=0.5]
        \coordinate (vl) at (0,0);
        \coordinate (v0) at (1.4,0);
        \coordinate (vr) at (2.8,0);
        \coordinate (v1) at (135:1);
        \coordinate (v2) at (180:0.8);
        \coordinate (v3) at (225:1);
        \coordinate (vcl) at (0.7,0);
        \coordinate (vcr) at (2.1,0);
        \draw[internal] (vl) -- (vcl);
        \draw[internal,fill=gray] (v0) circle (0.7);
        \draw[dummy] (vcr) -- (vr);
        \draw[internal] (v1) -- (vl);
        \draw[dummy] (v2) -- (vl);
        \draw[dummy] (v3) -- (vl);
        \fill (vl) circle (0.1);
        \fill (vcl) circle (0.1);
        \fill (vcr) circle (0.1);
      \end{tikzpicture}
    }}
  =
  \vcenter{\hbox{
      \begin{tikzpicture}[use Hobby shortcut, scale=0.5]
        \coordinate (vl) at (0,0);
        \coordinate (v0) at (1.4,0);
        \coordinate (vr) at (2.8,0);
        \coordinate (v1) at (135:1);
        \coordinate (v2) at (180:0.8);
        \coordinate (v3) at (225:1);
        \coordinate (vcl) at (0.7,0);
        \coordinate (vcr) at (2.1,0);
        \draw[internal] (vl) -- (vcl);
        \draw[internal,fill=gray] (v0) circle (0.7);
        \draw[dummy] (vcr) -- (vr);
        \draw[dummy] (v1) -- (vl);
        \draw[dummy] (v2) -- (vl);
        \draw[dummy] (v3) -- (vl);
        \fill (vl) circle (0.1);
        \fill (vcl) circle (0.1);
        \fill (vcr) circle (0.1);
      \end{tikzpicture}
    }}
  =0, 
\end{equation}
where dotted lines represent 
dummy field $x$. We use tilde to denote the quantities with tadpole contribution removed, and write
\begin{align}
	\beta_\Lambda & = \frac{1}{4!} \cdot \tilde \beta_{xxxx},
		      & \beta_a & = \frac{1}{3!} \cdot \tilde \beta_{axxx},
		      & \beta_{ab} &= \frac{1}{2} \cdot \tilde \beta_{abxx},
		      & \beta_{abc} & = \tilde \beta_{abcx}.
		      \label{eq:dummy}
\end{align}

The tensor structures, including the corresponding graphs and coefficients for all the considered RG functions, can be found in the form of supplementary \texttt{Mathematica} files.

\section{From general results to specific models\label{sec:examples}}
In this section we demonstrate the application of our general results to  particular scalar models. It is worth mentioning that we heavily rely on \texttt{FORM} \cite{Vermaseren:1992vn} to deal with index contractions and algebraic simplifications in the case of matrix fields. 

\subsection{Warming up with $O(n)$-symmetric model\label{sec:On}}

Our first example is the well-known $O(n)$ symmetric model, which has a long
history in the study of critical phenomena (see ref.~\cite{Kompaniets:2017yct}
and reference therein).
The following Euclidean Lagrangian describes the theory
\begin{align}
	\Lag = \frac{1}{2} \vec{\phi} (-\partial^2 + m^2) \vec{\phi}   
	+ \frac{\lambda}{4!} (\vec\phi \cdot \vec \phi)^2
	+ \frac{1}{2} g_{\phi\phi} d_{ab} \phi_a \phi_b,
	\label{eq:Lag_On}
\end{align}
where $\vec\phi = \{\phi_a\}$, $a=1,...,n$ is a $n$-component scalar field. We also add a quadratic operator involving traceless symmetric tensor $d_{ab}$ multiplied by a source $g_{\phi\phi}$. The anomalous dimension\footnote{In ref.~\cite{Kompaniets:2019zes} the notation $\gamma_{\mathcal{\tilde E}}=\gamma_{\phi\phi}$ is used.}  $\gamma_{\phi\phi}$ of the corresponding operator is related to the so-called crossover exponent (see, e.g., refs.~\cite{Kirkham:1981pu,Kompaniets:2019zes}) and can be found in our approach as 
\begin{align}
	\gamma_{\phi\phi} & = - \beta_{g_{\phi\phi}} + 2 \gamma_\phi
	\label{eq:crossover_exponent_On}
\end{align}
with $\beta_{g_{\phi\phi}}$ being the beta function of $g_{\phi\phi}$ and $\gamma_\phi$ corresponding to  the anomalous dimension of the field computed via eq.~\eqref{eq:phi_anom_dim}.
This and other RG functions can be easily obtained from our general result by means of substitutions 
\begin{align}
	\lambda_{abcd} & = \frac{\lambda}{3} \left( \delta_{ab} \delta_{cd} + \delta_{ac} \delta_{bd} + \delta_{ad} \delta_{bc} \right),
	\label{eq:lambda_On}
	\\
	m^2_{ab} & = m^2 \delta_{ab} + g_{\phi\phi} d_{ab}.
	\label{eq:m2_On}
\end{align}

In our calculation we find perfect agreement with previous computations.
Our new result is related to the six-loop contribution to the beta function of the vacuum energy $\beta_\Lambda$ for $g_{\phi\phi} = 0$ (see refs.~\cite{Kastening:1997ah,Larin:1997ek} for the five-loop expression).
Using the notation $g \equiv h \lambda$ (c.f.~ref.~\cite{Kastening:1997ah}) we have\footnote{In ref.~\cite{Kastening:1997ah} RG functions are defined as derivatives w.r.t $\ln \mu^2 = 2 \ln \mu$.
The factor $1\over2$ in \eqref{eq:beta_vac_On} is introduced for convenience.}  
\begin{align}
	\beta_v & \equiv \frac{1}{2} \cdot \frac{16 \pi^2}{m^4}  \beta_\Lambda 
		= 
		\frac{n}{4} 
		+ \frac{n(n+2)}{96} g^2 
	+ \frac{n(n+2) g^3}{1296}(n+8)(12 \zeta_3 - 25) \nonumber\\
		&+ 	
		\frac{ n (n+2)g^4}{82944} 
\Big[ 16 \zeta_3 (3 n^2 - 382 n - 1700)
	+ 96 \zeta_4 (4 n^2 + 39 n + 146) 
	 \nonumber\\
		& 
		\hspace{0.8cm}
		- 1024 \zeta_5(5n + 22) - 319 n^2 +  13968 n + 64864
\Big] \nonumber\\
		& +\frac{ n (n+2)g^5}{1866240} 
		\Big[
		 384 \zeta_3^2 (41 n^2 - 206 n - 888)
		-960 \zeta_6 (70 n^2 + 809 n + 2118)
			\nonumber \\
		& \hspace{0.8cm}
		+96 \zeta_5 (45 n^3  + 890 n^2 + 19348 n + 67440) 
		-288 \zeta_4 (n^3 + 294 n^2 + 3088 n + 9496)
		\nonumber \\
		& 
		\hspace{0.8cm}
		-48 \zeta_3(51 n^3 + 700 n^2 - 2964 n + 2024)
		- 1419 n^3 
		- 17124 n^2
		- 2166136 n 
		- 7308224 \nonumber\\
		& \hspace{0.8cm}
			+ 576 \zeta_7 (67 n^2 + 1405 n + 4306) 
			+ 1152 \zeta_3 \zeta_4 (2 n^2 + 145 n + 582)
	\Big].
	\label{eq:beta_vac_On}
\end{align}

By simple rescaling $\lambda \to 3! \lambda$, one can easily get the six-loop contributions to the RG functions for the Standard Model Higgs potential parameters (including the vacuum energy)  from the results of $O(4)$ theory. 

\subsection{Matrix models\label{sec:matrix}}

We consider matrix models with real and complex fields described by the following Lagrangians
\begin{align}
	\Lag = \frac{1}{2} \Tr \left[ \phi (-\partial^2 + \tau) \phi^T \right]  + \frac{\lambda_1}{4!} (\Tr\left[\phi \phi^T\right])^2 + \frac{\lambda_2}{4!} \Tr\left[\phi \phi^T \phi \phi^T\right] 
\label{eq:Lag_real_matrix}
\end{align}
for real $\phi$ and 
\begin{align}
	\Lag = \Tr[\phi (-\partial^2 + \tau) \phi^\dagger] +  \frac{\lambda_1}{4!}(\Tr[\phi \phi^\dagger])^2 + \frac{\lambda_2}{4!} \Tr[\phi \phi^\dagger \phi \phi^\dagger]
\label{eq:Lag_complex_matrix}
\end{align}
for complex $\phi$. 
To deal with matrix models we make use of the following decomposition (see also ref.~\cite{Litim:2020jvl})
\begin{align}
	\phi = \sum\limits_{a=1}^{N_a} \chi_a T_a,
	\label{eq:Phi_General_Decomposition}
\end{align}
where $\chi_a$ are real fields, and there are $N_a$ independent matrices $T_a$,  which encode all the degrees of freedom present in $\phi$. Substituting \eqref{eq:Phi_General_Decomposition} into either \eqref{eq:Lag_real_matrix} or \eqref{eq:Lag_complex_matrix}, we can rewrite the Lagrangians in the form \eqref{eq:Lag_general}. 
One can see that we completely get rid of the initial matrix indices of $\phi$ and replace them with a single one $a=1,\ldots,N_a$.  
Given eqs.~\eqref{eq:Lag_real_matrix} and \eqref{eq:Lag_complex_matrix}, for the fields $\chi^a$ to be canonically normalized, we have to ensure that
$\left[(T^a)^\dagger  \equiv \bar T^a\right]$
\begin{align}
	\Tr(T_a T_b^T) + \Tr(T_b T_a^T)  & = 2 \delta_{ab}, && \text{for real }\phi,
	\label{eq:T_norm_real}\\
	\Tr(T_a \bar T_b) + \Tr(T_b \bar T_a) & = \delta_{ab}, && \text{for complex }\phi.
	\label{eq:T_norm_complex}
\end{align}
As a consequence,  one can identify
\begin{align}
	m^2_{ab} & = -\tau \delta_{ab}, 
	\label{eq:m2_to_tau}\\
	\lambda_{abcd} & = \frac{\lambda_1}{4!} \left[
		T^{ab} T^{cd} + \text{perm.}
	\right]
	+ \frac{\lambda_2}{4!} 
	\left[
		T^{abcd} + \text{perm.}
	\right],
	\label{eq:lambda_to_l1_l2}
\end{align}
where 
\begin{align}
	T^{ab} 
	& = \Tr( T_a T_b^T) 
	\equiv 
	\TrR{a}{b}
	, & T^{abcd} & = \Tr( T_a T_b^T T_c T_d^T)
	\equiv
	\TrR{a}{b}{c}{d}
	, & &\text{for real } \phi, \\ 
	T^{ab} 
	& = \Tr( T_a \bar T_b)
	\equiv 
	\TrC{a}{b}
	, & T^{abcd} & = \Tr( T_a \bar T_b T_c \bar T_d)
	\equiv
	\TrC{a}{b}{c}{d}
	, & &\text{for complex } \phi.
\end{align}

In eq.~\eqref{eq:lambda_to_l1_l2} all 24 permutations of the indices $abcd$ are taken into account. Obviously, the number of terms can be reduced in specific models. 
In the following subsections we provide some details of our calculations for the cases discussed in the literature.  

\subsubsection{Real anti-symmetric field\label{sec:On_Asym}}

The Lagrangian of the model is given by eq.~\eqref{eq:Lag_real_matrix} with 
$\phi$ being an antisymmetric $n\times n$ matrix, $\phi^T = -\phi$. The 
model was considered in refs.~\cite{Antonov_2013,Antonov:2017pqv} and the four-loop results can be found in ref.~\cite{lebedev2018critical}. 

To use our general formulae, we utilize the decomposition \eqref{eq:Phi_General_Decomposition} with $N_a = \frac{n(n-1)}{2}$ and $T_a = t_a$ corresponding to antisymmetric generators of $SO(n)$. The latter satisfy
\begin{align}
		\Tr (t^a t^b)  = T_f \delta_{ab}, \quad
		t^a_{ij} t^a_{kl}  = \frac{T_f}{2} \left( \delta_{il} \delta_{jk} - \delta_{ik} \delta_{jl}\right).
\label{eq:Algebra_On_Asym}
\end{align}
To keep the standard normalization for the fields $\chi_a$, we use $T_f = 1$ (see eq.~\eqref{eq:T_norm_real}). 
The number of terms in eq.~\eqref{eq:lambda_to_l1_l2} can be reduced ($\Tr t^a t^b \ldots \equiv  T^{ab\ldots}$)
\begin{align}
	\lambda_{abcd} & = \frac{\lambda_1}{3} 
	\left[ 
		  T^{ab} T^{cd} 
		+ T^{ac} T^{bd} 
		+ T^{ad} T^{bc} 
	\right] 
		       + \frac{\lambda_2}{3} 
	\left[
		    T^{abcd}
		  + T^{abdc} 
		  + T^{acbd} 
	\right],
	\label{eq:Lambda_On_Asym}
\end{align}
where we used the cyclic symmetry of the trace operation and the fact that $t_a^T = - t_a$. 

By means of eq.~\eqref{eq:Algebra_On_Asym} we write down the rules, which allow one to simplify the products of traces involving $t_a$ with some of the indices contracted.  
Substituting \eqref{eq:Lambda_On_Asym} into the general expression for $\beta_{abcd}$, and performing the above-mentioned algebraic simplifications, we obtain 
$\beta_{abcd}$ of the form
\begin{align}
	4! \beta_{abcd} & = f_1(\lambda_1,\lambda_2,n) \left[
		T^{ab}
		T^{cd}
		+ \text{perm.}
	\right] 
	+ f_2 (\lambda_1, \lambda_2, n) \left[
		T^{abcd}
+ \text{perm.}
	\right],
\label{eq:beta_example}
\end{align}
where $f_{1,2}(\lambda_1, \lambda_2,n)$ are some polynomials of their arguments. It is possible to extract 
the beta functions for $\lambda_1$ and $\lambda_2$ from eq.~\eqref{eq:beta_example} by applying suitable projectors.  
However, one can also use the fact that by construction $\beta_{abcd}$ is symmetric in all the indices. Setting the latter equal to each other in the end of calculation, we have
\begin{align}
	\beta_{aaaa} & = f_1(\lambda_1,\lambda_2,n) 
	[T^{aa}]^2
	+ f_2 (\lambda_1, \lambda_2, n) 
	[T^{aaaa}]
		, 
	\quad \text{no sum over } a.
\label{eq:beta_example_2}
\end{align}
	Comparing eqs.~\eqref{eq:beta_example_2} and 
	\eqref{eq:Lambda_On_Asym} with $b=c=d=a$, one can easily deduce that
\begin{align}
	\beta_{\lambda_1}& = f_1(\lambda_1, \lambda_2,n) , \quad \beta_{\lambda_2} = f_2(\lambda_1, \lambda_2,n).
	\label{eq:beta_wo_projectors}
\end{align}
We utilize this approach to obtain relevant RG functions up to the six-loop level. 
Our results agree with that given in refs.~\cite{Antonov:2017pqv,lebedev2018critical}
\footnote{Note that in ref.~\cite{Antonov:2017pqv} the notation $\lambda_i = g_i$ is used and the RG functions are expanded in $g_i/(8 \pi^2)$.}. 
It is worth noting that for $n=2$ and $n=3$ the model is equivalent to one-component $\phi^4$ and the $O(3)$-vector theory considered in sec.~\ref{sec:On}, respectively.
Indeed, combining $\lambda = \lambda_1 + {1\over 2} \lambda_2$ and computing $\beta_\lambda = \beta_{\lambda_1} + {1 \over 2} \beta_{\lambda_2}$ for $n=2$ and $n=3$ we get the expected results.  

Full six-loop beta functions and anomalous dimensions are available online as supplementary material. 
For convenience, we present here our expressions for the one-loop 
\begin{align}
	\beta^{(1)}_{\lambda_1} & =
	\frac{\lambda_1^2}{6} (n^2 - n + 16)
		+
		\frac{\lambda_1 \lambda_2}{3}
		(2n -1)
		+ \frac{\lambda_2^2}{2}, \\
	\beta^{(1)}_{\lambda_2} & =
		4 \lambda_1 \lambda_2 
		+ \frac{\lambda_2^2}{6} (2n - 1), \\
	\gamma^{(1)}_\tau & \equiv - \beta^{(1)}_{\tau}/\tau  = -  
			\frac{\lambda_1}{6} (n^2 - n + 4)
			-\frac{\lambda_2}{6} (2 n -1)
	,
	\label{eq:rg_On_Asym_1l}
\end{align}
	and two-loop RG functions 
\begin{align}
	\gamma^{(2)}_\phi & = 
	\frac{1}{288}
	\left[
			4 \lambda_1^2
		+ \lambda_2^2\right]
(n^2 - n + 4)
+ \frac{\lambda_1\lambda_2}{36} (2n-1),
\\
	\beta^{(2)}_{\lambda_1} & =
	\frac{\lambda_1^3}{6} (3n - 3n^2 - 28)
	- \left[
		\frac{11\lambda_1^2 \lambda_2}{9}
		+ \frac{\lambda_2^3}{12}
	\right]
	(2n-1)
	- \frac{\lambda_1 \lambda^2_2}{72}
		(5 n^2 - 5n + 164),
		\\
	\beta^{(2)}_{\lambda_2} & =
	\frac{\lambda_1^2 \lambda_2}{18}
	(5n - 5n^2 - 164)
	- \frac{11\lambda_1 \lambda_2^2}{9}
	(2n - 1)
	+ \frac{\lambda_2^3}{24} (n -n^2 - 20), \\
	\gamma^{(2)}_\tau &  =   
	\frac{5}{144}
	\left[
			4 \lambda_1^2
		+ \lambda_2^2\right]
(n^2 - n + 4)
+ \frac{5\lambda_1\lambda_2}{18} (2n-1).
	\label{eq:rg_On_Asym_2l}
\end{align}

\subsubsection{$O(n) \times O(m)$ model\label{sec:OnOm}}

Let us now consider a matrix model, which is invariant under $O(n)\times O(m)$ group.
It describes the critical thermodynamics of frustrated spin systems with noncollinear and noncoplanar ordering (see, e.g., ref.~\cite{Kompaniets:2019xez} and references therein).  In refs.~\cite{Calabrese:2003ww} five-loop results are presented in terms of $u = \lambda_1 + \lambda_2$, and $v=\lambda_2$. Six-loop RG functions are also known \cite{Kompaniets:2019xez} in terms of $g_i=\lambda_i$.

The Landau-Wilson Lagrangian can be written in the form \eqref{eq:Lag_real_matrix} with 
$\phi = \{\phi_{\alpha i}\}$ being $n\times m$ real matrix field, and $\alpha = 1,\ldots,n$, $i=1,\ldots,m$.

To compute relevant RG functions from our general result we interpret $\chi_a$ in eq.~\eqref{eq:Phi_General_Decomposition} as $N_a = n\cdot m$ matrix elements of $\phi$, so that
each of $n\times m$ real matrices $T^a$ has only one non-zero element
\begin{align}
	(T_a)_{\alpha i} = \sqrt{T_f} \cdot \delta_{\alpha, ((a-1)~\text{div}~m)+1} 
			   \delta_{i,((a-1)~\text{mod}~m)+1},
			   \label{eq:T_OnOm}
\end{align}
where we introduce $T_f = 1$ for convenience. As a consequence\footnote{ Notice here that $T_a T_b^T$ are $n\times n$ matrices, while $T_a^T T_b$ have $m\times m$ dimension.}, we have
\begin{align}
	\Tr( T_a T^T_b ) = \Tr(T^T_b T_a) = T_f \delta_{ab}, \quad T_{\alpha i}^a T_{\beta j}^a = T_f \delta_{\alpha \beta} \delta_{ij}.
\label{eq:Algebra_OnOm}
\end{align}
The quartic self-coupling is given by
\begin{align}
	\lambda_{abcd} 
	& = \frac{\lambda_1}{3} 
	\left[ 
		  \TrR{a}{b} \TrR{c}{d}
		+ \TrR{a}{c} \TrR{b}{d}
		+ \TrR{a}{d} \TrR{b}{c}
	\right] \nonumber \\
	& + \frac{\lambda_2}{6} 
	\left[
		  \TrR{a}{b}{c}{d}
		+ \TrR{a}{b}{d}{c}
		+ \TrR{a}{c}{b}{d}
		+ \TrR{a}{c}{d}{b}
		+ \TrR{a}{d}{b}{c}
		+ \TrR{a}{d}{c}{b}
	\right],
\label{eq:Lambda_OnOm}
\end{align}
	where to reduce the number of terms in LHS, we use the fact that
\begin{align*}
	\Tr(A) = \Tr(A^T), \quad (T_a T^T_b T_c  T^T_d)^T = T_d T^T_c T_b T^T_a  
\end{align*}
so 
\begin{align}
	\TrR{a}{b}{c}{d} \equiv \Tr(T_a T^T_b T_c  T^T_d)
& =  
\Tr(T_d T^T_c T_b T^T_a)
\equiv \TrR{d}{c}{b}{a}
.
\label{eq:Tr_sym_OnOm}
\end{align}
To extract the RG functions, we substitute \eqref{eq:Lambda_OnOm} together with \eqref{eq:m2_to_tau} into $\beta_{abcd}$, $\beta_{ab}$ and $\gamma_{ab}$ and use the rules \eqref{eq:Algebra_OnOm} to simplify the products of traces involving $T_a$ and $T_b^T$.

We use known results \cite{Calabrese:2003ww, Kompaniets:2019xez} to cross-check our expressions, which at the one-loop order are given by
\begin{align}
	\beta^{(1)}_{\lambda_1} & =
	\frac{\lambda_1^2}{3} (8 + n m)
		+
		\frac{2 \lambda_1 \lambda_2}{3}
		(1 + n + m)
		+ \lambda_2^2, \\
	\beta^{(1)}_{\lambda_2} & =
		4 \lambda_1 \lambda_2 
		+ \frac{\lambda_2^2}{3} (4 + m + n), \\
	\gamma^{(1)}_\tau & \equiv - \beta^{(1)}_{\tau}/\tau	
	= - \frac{\lambda_1}{3} ( 2 + n m)
			-\frac{\lambda_2}{3} (1 + m + n),
	\label{eq:rg_OnOm_1l}
\end{align}
	while at two loops we have
\begin{align}
	\gamma^{(2)}_\phi & = 
	\frac{\lambda_1^2}{36}
	(2 + m n)
	+\frac{\lambda_1 \lambda_2}{18}
	(1 + m + n )
	+\frac{\lambda^2_2}{72}
	(3 + m n + m  + n),
	\\
	\beta^{(2)}_{\lambda_1} & =
	-\frac{\lambda_1^3}{3} (14 + 3 m n)	
	- \frac{22 \lambda_1^2 \lambda_2}{9}
	(1 + m +n)
	\nonumber\\
				&- \frac{\lambda_1 \lambda_2^2}{18}
		(87 + 5 (m n +  m +  n))
	- \frac{\lambda^2_3}{3}
		(3 + m + n),
		\\
	\beta^{(2)}_{\lambda_2} & =
	-\frac{\lambda_1^2 \lambda_2}{9}
	(82 + 5 m n )
	- \frac{2\lambda_1 \lambda_2^2}{9}
	(11 (n+m) + 29)
	- \frac{\lambda_2^3}{6} ( 17 + mn + 3 (m+n) ), \\
	\gamma^{(2)}_\tau &  =   
	\frac{5\lambda_1^2}{18}
	(2 + m n)
	+\frac{5\lambda_1 \lambda_2}{9}
	(1 + m + n)
	+\frac{5\lambda^2_2}{36}
	(3 + m n + m  + n).
	\label{eq:rg_OnOm_2l}
\end{align}
In addition, we extend to the six-loop order the anomalous dimensions of quadratic operators considered in refs.~\cite{DePrato:2006jx,Pelissetto:2007gw}:
\begin{align}
	Q^{(1)}_{\alpha i \beta j} & = \phi_{\alpha i} \phi_{\beta j} - \phi_{\alpha j} \phi_{\beta i}, \\
	Q^{(2)}_{\alpha i \beta j} & = \frac{1}{2} \left(
		  \phi_{\alpha i} \phi_{\beta j} 
		+ \phi_{\alpha j} \phi_{\beta i}
	\right)
	- \frac{1}{n} \delta_{\alpha \beta}  \phi_{\delta i} \phi_{\delta j}
	- \frac{1}{m} \delta_{i j} \phi_{\alpha k} \phi_{\beta k} 
	+ \frac{1}{n m} \delta_{\alpha \beta} \delta_{i j} 
	\phi_{\delta k} \phi_{\delta k}, \\
		Q^{(3)}_{i j} & = \phi_{\delta i} \phi_{\delta j} - \frac{1}{m} \delta_{ij} \phi_{\delta k} \phi_{\delta k} \equiv \tilde Q^{(3)}_{\alpha \beta i j} \delta_{\alpha \beta}, \\
		Q^{(4)}_{\alpha \beta} & = \phi_{\alpha k} \phi_{\beta k} - \frac{1}{n} \delta_{\alpha \beta} \phi_{\delta k} \phi_{\delta k} \equiv \tilde Q^{(4)}_{\alpha \beta i j} \delta_{i j}
		\label{eq:quadratic_perturbations_OnOm}
\end{align}
	and belonging to different representations of $O(n)\times O(m)$. 
	The operators can be 
	treated in our approach in a similar fashion. 
	We assume that the perturbations can be added to the Lagrangian with the corresponding sources (``masses'') and rewritten in terms of $\chi$-fields as, e.g.,
\begin{align}
	\left[ \tilde m^2_{1,\alpha i \beta j}\right]
	\left[ Q^{(1)}_{\alpha i \beta j}\right] 
	& = \left[\tilde m^{2}_{1,cd}
	T^c_{\alpha i} T^d_{\beta j} \right]
	\left[
	\chi^a \chi^b \left( T^a_{\alpha i} T^b_{\beta j} - T^a_{\alpha j} T^b_{\beta j} \right)
\right] \equiv m^2_{1,ab} \chi^a \chi^b 
	\\
	m^2_{1,ab}	& = \frac{1}{2} 
		\tilde m^{2}_{1,cd} 
		\left[ \TrR{a}{c} \TrR{b}{d} 
		- 
		\TrR{a}{d}{b}{c}
	+ (a\leftrightarrow b)
	\right]
.
\end{align}
Since the operators \eqref{eq:quadratic_perturbations_OnOm} do not mix under renormalization, we use the following substitutions\footnote{Given eq.~\eqref{eq:Tr_sym_OnOm}, one can prove that tensors multiplying $\tilde m^2_{i,cd}$  are symmetric in $a\leftrightarrow b$ and $c\leftrightarrow d$.}
\begin{align}
	Q^{(1)}_{\alpha i \beta j}:\quad  m^2_{ab} & \Rightarrow 
	\frac{\tilde m^2_{1,cd}}{2} 
	\left[ 
		\TrR{a}{c} \TrR{b}{d}
		- 
		\TrR{a}{d}{b}{c} 
		+ ( a \leftrightarrow b)
	\right],
	\\
	Q^{(2)}_{\alpha i \beta j}:\quad  m^2_{ab} & \Rightarrow 
	\frac{\tilde m^2_{2,cd}}{4} 
	\left[ 
		\TrR{a}{c}\TrR{b}{d}
		+
		\TrR{a}{d}{b}{c}
		- \frac{2}{n} \TrR{a}{c}{d}{b} 
	\right.
				       \nonumber\\
						   &  
		 \phantom{\Rightarrow\frac{\tilde m^2_{2,cd}}{4} }
						   \left.
		- \frac{2}{m} 
		\TrR{a}{b}{c}{d} 
		+ (a\leftrightarrow b)
	\right] + \frac{\tilde m^2_{2,cd}}{n m} 
	\TrR{a}{b} \TrR{c}{d} 
	,
	\\
	Q^{(3)}_{i j}:\quad  m^2_{ab} & \Rightarrow 
\tilde m^2_{3,cd}
\left[
	\frac{1}{2}
	\left[ 
		\TrR{a}{c}{d}{b} 
		+ (a\leftrightarrow b)
	\right]
	- \frac{1}{m} 
	\TrR{a}{b} \TrR{c}{d} 
\right]
	,
	\\
	Q^{(4)}_{\alpha \beta}:\quad  m^2_{ab} & \Rightarrow 
\tilde m^2_{4,cd}
\left[
	\frac{1}{2}
	\left[ 
		\TrR{a}{b}{c}{d} 
		+ (a\leftrightarrow b)
	\right]
	- \frac{1}{n} 
	\TrR{a}{b} \TrR{c}{d}
\right]
\end{align}
and extract the beta functions $(\beta_{\tilde m^2_i})_{cd} \equiv -\tilde \gamma_{i} \cdot \tilde m^2_{i,cd}$ of $\tilde m^2_{i,cd}$, $i=1,4$
from the corresponding terms in $\beta_{ab}$. The RG functions for the operators $Q^{(i)}$ \eqref{eq:quadratic_perturbations_OnOm} are obtained by adding the contribution from the field anomalous dimension $\gamma_\chi=\gamma_\phi$:
\begin{align}
	\gamma_{Q_i} = \tilde \gamma_{i} {+} 2 \gamma_\phi.
\end{align}
A welcome check of the result is the fact that for $v=0$ all $\gamma_{Q_i}$ coincide. We also compare our expressions with that given in ref.~\cite{Pelissetto:2007gw} and find perfect agreement up to five loops\footnote{The results of ref.~\cite{Pelissetto:2007gw} are written in terms of $(u,v)/(8 \pi^2)$.}.
We present here our one-loop, 
\begin{align}
	\gamma^{(1)}_{Q_1} & = -\frac{2 \lambda_1 - \lambda_2}{3},
		\\
		\gamma^{(1)}_{Q_2} & = -\frac{2 \lambda_1 + \lambda_2}{3},
		\\
		\gamma^{(1)}_{Q_3} & = -\frac{2 \lambda_1 + (1 + n) \lambda_2 }{3}, 
		\\
		\gamma^{(1)}_{Q_4} & = -\frac{2 \lambda_1 + (1 + m) \lambda_2 }{3}, 
	\label{eq:rg_OnOm_oper_1l}
\end{align}
and two-loop results
\begin{align}
	\gamma^{(2)}_{Q_1} & = 
	\frac{\lambda_1^2}{9} (6 + m n)
	+ \frac{2\lambda_1 \lambda_2}{9} (1 + m + n)
	+ \frac{\lambda^2_2}{18} (1-m - n),
		\\
		\gamma^{(2)}_{Q_2} & = 
	\frac{\lambda_1^2}{9} (6 + m n)
	+ \frac{2\lambda_1 \lambda_2}{9} (3 + m + n)
	+ \frac{\lambda^2_2}{18} (9 + m + n),
		\\
		\gamma^{(2)}_{Q_3} & = 
	\frac{\lambda_1^2}{9} (6 + m n)
	+ \frac{2\lambda_1 \lambda_2}{9} (3 + m + 3 n)
	+ \frac{\lambda^2_2}{18} (9 + m n + m  + 3n),
		\\
		\gamma^{(2)}_{Q_4} & = 
	\frac{\lambda_1^2}{9} (6 + m n)
	+ \frac{2\lambda_1 \lambda_2}{9} (3 + 3m +  n)
	+ \frac{\lambda^2_2}{18} (9 + m n + 3m  + n).
	\label{eq:rg_OnOm_oper_2l}
\end{align}
The six-loop expressions are available as supplementary material.

\subsubsection{Complex anti-symmetric field \label{sec:Un_Asym}}

Let us now generalize the model discussed in sec.~\ref{sec:On_Asym} and consider complex antisymmetric $n\times n$ matrices  $\phi$.  The corresponding Lagrangian \eqref{eq:Lag_complex_matrix} can be used to study phase transitions in quantum Fermi systems within the RG approach (see ref.~\cite{komarova2013temperature}). We decompose the field via \eqref{eq:Phi_General_Decomposition} with $N_a = n(n-1)$ and antisymmetric 
\begin{align*}
	T^a = 
	\begin{cases}
		  t^a & a = 1,\ldots,n(n-1)/2, \\
		  i t^{a} & a = 1+n(n-1)/2,\ldots,n(n-1).
	\end{cases}
\end{align*}
The latter are written in terms of generators $t^a$ of $SO(n)$.  Given $\Tr (t^a t^b) = T_f \delta_{ab}$, one can derive $\left[(T^a)^\dagger  \equiv \bar T^a\right]$
\begin{align}
		&\Tr (T^a \bar T^b) + \Tr (T^b \bar T^a) = -2 T_f \delta_{ab}, \quad 
		 T^a_{ij} T^a_{kl}  = 
		\sum\limits_{b=1}^{N_a/2} \left( t^b_{ij} t^b_{ij} + i^2 t^b_{ij} t^b_{ij}\right) = 0,\nonumber \\
		& T^a_{ij} \bar T^a_{kl}   = 
		\sum\limits_{b}^{N_a/2}
		\left(t_{ij}^b (-t^b_{kl})
		+ i t_{ij}^b (+i t^b_{kl})\right) 
		= - T_f \left( \delta_{il} \delta_{jk} - \delta_{ik} \delta_{jl}\right).
\label{eq:Algebra_Un_Asym}
\end{align}
One can see from eq.~\eqref{eq:T_norm_complex} that for $T_f = -1/2$ the fields $\chi^a$ are canonically normalized. The self-coupling \eqref{eq:lambda_to_l1_l2} is given by 
\begin{align}
	\lambda_{abcd} & = \frac{\lambda_1}{12} 
	\left[ 
		\TrC{a}{b}
		\TrC{c}{d}
		+ \text{11 permutations}
	\right] \nonumber\\
		       & + \frac{\lambda_2}{6} 
	\left[
		\TrC{a}{b}{c}{d} 
		+ \TrC{a}{b}{d}{c} 
		+ \TrC{a}{c}{b}{d} 
             \TrC{c}{a}{d}{b} 
		+ \TrC{b}{a}{c}{d} 
		+ \TrC{b}{a}{d}{c} 
	\right].
	\label{eq:Lambda_Un_Asym}
\end{align}
	In writing the latter we take into account that 	
\begin{align*}
	\Tr(A) = \Tr(A^T), \quad (T^a \bar T^b T^c \bar T^d)^T = (T^d)^* (T^c)^T (T^b)^* (T^a)^T = 
	\bar T^d T^c \bar T^b T^a,
\end{align*}
so, e.g.,
\begin{align*}
	\TrC{a}{b}{c}{d} \equiv
\Tr(T^a \bar T^b T^c \bar T^d)
& =  \Tr( \bar T^d T^c \bar T^b T^a) = 
\Tr(T^a \bar T^d T^c \bar T^b)
\equiv
\TrC{a}{d}{c}{b}
.
\end{align*}

The expressions for the RG functions are available in literature up to the five-loop level\footnote{In terms of $g_i = \lambda_i/6$.} \cite{Kalagov:2015gra}. We extend these results up to six loops. The one-loop contributions read 
\begin{align}
	\beta^{(1)}_{\lambda_1} & =
	\frac{\lambda_1^2}{12} (n^2 - n + 8)
		+
		\frac{\lambda_1 \lambda_2}{3}
		(n -1)
		+ \frac{\lambda_2^2}{4}, \\
	\beta^{(1)}_{\lambda_2} & =
		\lambda_1 \lambda_2 
		+ \frac{\lambda_2^2}{12} (2n - 5), \\
	\gamma^{(1)}_\tau & \equiv - \beta^{(1)}_{\tau}/\tau  = 
	-  \frac{\lambda_1}{12} (n^2 - n + 2)
			-\frac{\lambda_2}{6} (n -1)
	,
	\label{eq:rg_Un_Asym_1l}
\end{align}
	while two-loop  corrections are
\begin{align}
	\gamma^{(2)}_\phi & =
	\frac{\lambda_1^2}{576} (n^2 - n - 2)
	+\frac{\lambda_1 \lambda_2}{144} ( n - 1)
	+\frac{\lambda^2_2}{1152} (n^2 - 3n +4),
	\\
	\beta^{(2)}_{\lambda_1} & =
	\frac{\lambda_1^3}{48} (3n - 3n^2 - 14)
		-
		\frac{11\lambda^2_1 \lambda_2}{36}
		(n -1)
		+
		\frac{\lambda_1 \lambda^2_2}{288}
		(15 n - 5n^2 - 92)
		- \frac{\lambda_2^3}{24} (n-2), \\
	\beta^{(2)}_{\lambda_2} & =
	\frac{\lambda_1^2 \lambda_2}{144} (5n - 5n^2 - 82)
		+
		\frac{\lambda_1 \lambda^2_2}{36}
		(20 - 11 n)
		+
		\frac{\lambda^3_2}{96}
		(7 n - n^2 - 20),
		\\
	\gamma^{(2)}_\tau & =
	\frac{5\lambda_1^2}{288} (n^2 - n - 2)
	+\frac{5\lambda_1 \lambda_2}{72} ( n - 1)
	+\frac{5\lambda^2_2}{576} (n^2 - 3n +4).
	\label{eq:rg_Un_Asym_2l}
\end{align}

All results at six loops are available online as supplementary material.

\subsubsection{$U(n) \times U(m)$ model}
Consider now eq.~\eqref{eq:Lag_complex_matrix} with general complex $n\times m$ matrix field $\phi = \{\phi_{\alpha i}\}$. 
The model can be used to study phase transitions in massless QCD and five-loop RG functions are available in literature~\cite{Calabrese:2004uk}. We compute the six-loop contributions by means of decomposition \eqref{eq:Phi_General_Decomposition} with $N_a = 2 n m$ and
$T^a$ being complex $n\times m$ matrices  (c.f. eq.~\eqref{eq:T_OnOm})
\begin{align}
	(T_a)_{\alpha i} = \sqrt{T_f}
	\begin{cases}
		\delta_{\alpha, ((a-1)~\text{div}~m)+1} 
		\delta_{i,((a-1)~\text{mod}~m)+1}, &  a = 1,\ldots m n \\
		i T^{a - m n}_{\alpha i}, & a = m n + 1,\ldots 2 m n, 
		   \end{cases}
		   \label{eq:generators_Un_Um}
\end{align}
	satisfying
\begin{align*}
	\Tr( T_a \bar T_b )  +   
	\Tr( T_b \bar T_a )  = 2 T_f \delta^{ab},
\end{align*}
and
\begin{align}
		T^a_{\alpha j} T^a_{\beta k}  & = 0, 
					      & 
		\bar T^a_{j \alpha} \bar T^a_{k \beta} &  
		= 0, 
						       & T^a_{\alpha j} \bar T^a_{k \beta} & 
						       = 2 T_f \delta_{\alpha\beta} \delta_{jk}. 
						       \label{eq:Algebra_UnUm}
\end{align}
The calculations are carried out with $T_f = 1/2$ and the following representation of general self-coupling
\begin{align}
	\lambda_{abcd} & = 
	\frac{\lambda_1}{12} \left[ 
		\TrC{a}{b} \TrC{c}{d}
	+ \text{11 perms} \right]
	+ \frac{\lambda_2}{12} \left[ 
	\TrC{a}{b}{c}{d} 
+ \text{11 perms}\right],
	\label{eq:Lambda_UnUm}
\end{align}
where among all 24 permutations we exclude only those that correspond to the swapping between pairs of indices. 

Our calculation employs eq.~\eqref{eq:Algebra_UnUm} and renders at one loop
\begin{align}
	\beta^{(1)}_{\lambda_1} & =
	\frac{\lambda_1^2}{6} (4 + n m)
		+
		\frac{\lambda_1 \lambda_2}{3}
		(n + m)
		+ \frac{\lambda_2^2}{2}, \\
	\beta^{(1)}_{\lambda_2} & =
		\lambda_1 \lambda_2 
		+ \frac{\lambda_2^2}{6} (n + m), \\
	\gamma^{(1)}_\tau & \equiv - \beta^{(1)}_{\tau}/\tau  = 
	-  \frac{\lambda_1}{6} (1 + n m)
			-\frac{\lambda_2}{6} (n + m)
			,
	\label{eq:rg_UnUm_1l}
\end{align}
and at two loops 
\begin{align}
	\gamma^{(2)}_\phi & = 
	\frac{1}{288}(\lambda_1^2 + \lambda_2^2)(1 + m n)	
	+\frac{\lambda_1 \lambda_2}{144} (n + m),\\
	\beta^{(2)}_{\lambda_1} & =
	- \frac{\lambda_1^3}{24}
 (7 + 3 m n)	
	- \frac{11 \lambda_1^2 \lambda_2}{36}
	(m +n)
		- \frac{\lambda_1 \lambda_2^2}{72}
		(41 + 5 m n)
	- \frac{\lambda^2_3}{12}
		(m + n),
		\\
	\beta^{(2)}_{\lambda_2} & =
	- \frac{\lambda_1^2\lambda_2}{72}
 (41 + 5 m n)	
	- \frac{11 \lambda_1 \lambda_2^2}{36}
	(m +n)
	- \frac{\lambda^2_3}{24}
		(5 + m n),
	\\
	\gamma^{(2)}_\tau & = 
	\frac{5}{144}(\lambda_1^2 + \lambda_2^2)(1 + m n)	
	+\frac{5\lambda_1 \lambda_2}{72} (n + m)	
	.
	\label{eq:rg_UnUm_2l}
\end{align}
The full results are available as supplementary material.
It is worth noting that for $m=n$ we get the four-loop results obtained in ref.~\cite{Steudtner:2020tzo} for the case of $U(n) \times U(n)$ model.\footnote{In ref.~\cite{Steudtner:2020tzo} the RG functions are written in terms of $a_u \equiv h n \lambda_2/24$ and $a_v \equiv h n^2 \lambda_1/24$.}

\subsubsection{Field in the adjoint representation of $SU(n)$ \label{sec:adj-scalar2}}
In recent ref.~\cite{Hnatic:2020kyo} a model 
with $\phi$ being hermitian matrix field in the adjoint representation of $SU(n)$ is analyzed  both with perturbative and non-perturbative methods. 
In addition, the model was also considered as an example of application of the \texttt{ARGES} code~\cite{Steudtner:2021fzs}.
We generalize the Lagrangian of ref.~\cite{Hnatic:2020kyo} and include also a cubic term\footnote{The term breaks $Z_2$ symmetry $\phi \to -\phi$ imposed in ref.~\cite{Hnatic:2020kyo}.} (see also refs.~\cite{Ruegg:1980gf,Murphy:1983rf,Litim:2020jvl}) together with the vacuum energy (we rescale $f$ and $\lambda_2$ for convenience)
\begin{align}
	\Lag = \frac{1}{2} \Tr \left[ \phi (-\partial^2 + m^2) \phi\right]
	+ \frac{\sqrt{n} f}{3!} \Tr \phi^3 + \frac{\lambda_1}{4} \left(\Tr \phi^2\right)^2 +
	\frac{n \lambda_2}{4!} \Tr\phi^4 + \Lambda.
	\label{eq:Lag_SUN}
\end{align}
 Obviously, we can easily treat the model 
in our approach by means of the decomposition \eqref{eq:Phi_General_Decomposition} with $T_a$ being $SU(n)$ generators.  The latter satisfy the well-known relations \cite{Cvitanovic:1976am}
\begin{align}
	\Tr(T_a T_b) = T_f \delta_{ab}, \quad T^a_{ij} T^a_{kl} = T_f \left(\delta_{il} \delta_{kl} - \frac{1}{n} \delta_{ij} \delta_{kl}\right).
	\label{eq:Algebra_SUn}
\end{align}
We utilize the normalization $T_f = 1$ and substitute 

\begin{align}
	\lambda_{abcd} 
	& = \frac{\lambda_1}{3} 
	\left[ 
		  T^{ab} T^{cd}
		+ T^{ac} T^{bd}
		+ T^{ad} T^{bc}
	\right] 
	\nonumber \\
	& \hspace{0.33cm}+ \frac{n \lambda_2}{6} 
	 \left[
		  T^{abcd}
		+ T^{abdc}
		+ T^{acbd}
		  + T^{acdb}
		  + T^{adbc}
		  + T^{adcb}
	\right]
\label{eq:Lambda_SUn}
\end{align}
	together with
\begin{align}
	h_{abc} = \frac{\sqrt{n} f}{2} \left[ T^{abc} + T^{bac}\right]. 
	\label{eq:tri_SUn}
\end{align}
	We obtain the RG functions up to the six-loop level, and at one loop we have
\begin{align}
	\beta^{(1)}_{\lambda_1} & =
	\frac{\lambda_1^2}{3} (7 + n^2)
		+
		\frac{\lambda_1 \lambda_2}{3}
		(4n^2 -6)
		+ \lambda_2^2(3 + n^2), \\
	\beta^{(1)}_{\lambda_2} & =
		4 \lambda_1 \lambda_2 
		+ \frac{\lambda_2^2}{3} (2n^2 - 18), \\
	\beta^{(1)}_{f} & = f \left[ 2 \lambda_1  + \lambda_2 (n^2 - 6)\right], 
	\\
	\beta^{(1)}_{m^2} & = 
	\frac{m^2}{3} \left[ 
		\lambda_1(n^2 + 1)
		+ \lambda_2 (2 n^2 - 3)
	\right]  + \frac{f^2}{2} (n^2 - 4), \\
	\beta^{(1)}_{\Lambda} & = 
	\frac{m^4}{2} (n^2 - 1).
	\label{eq:rg_SUn_1l}
\end{align}
	The two-loop expressions are given by 
\begin{align}
	\gamma^{(2)}_{\phi} & = 
 	\frac{\lambda_1^2}{36} (1 + n^2)
 	-\frac{\lambda_1 \lambda_2}{18} (3 - 2n^2)
 	+\frac{\lambda_2^2}{72} (18 - 6 n^2 + n^4),
	\\
	\beta^{(2)}_{\lambda_1} & =
	-\frac{\lambda_1^3}{3} (11 + 3 n^2)
	+\frac{22\lambda_1^2 \lambda_2}{9} (3 - 2 n^2)
	\nonumber\\
				&
	-\frac{\lambda_1\lambda^2_2}{18} (306 + 42 n^2 + 5 n^4)
	+ \frac{\lambda_2^3}{3}(36 + 9 n^2 - 2n^4), \\
	\beta^{(2)}_{\lambda_2} & =
	\frac{\lambda_1^2 \lambda_2}{9} (77 + 5n^2)
	+\frac{2\lambda_1\lambda^2_2}{9} (123 - 22 n^2 )
	- \frac{\lambda_2^3}{6}(174 - 16 n^2 + n^4), \\
	\beta^{(2)}_{f} & = \frac{f}{24} 
	\left[ 
	4 \lambda_1 \lambda_2 (105 - 22 n^2)
	- 2 \lambda_1^2  (35 + 3 n^2)
	- \lambda_2^2 (630 -138 n^2 +11 n^4)
\right]
	\\
	\beta^{(2)}_{m^2} & = 
	\frac{5m^2}{36} \left[ 
		4 \lambda_1 \lambda_2 (3 - 2 n^2)
		- 2 \lambda_1^2 (1 + n^2)
		- \lambda_2^2 (18 - 6 n^2 + n^4)
	\right]  
	\nonumber\\
			  & + \frac{f^2}{12} 
	\left[
	\lambda_1 (36 - 5 n^2 - n^4)
	-3 \lambda_2 (36 - 17 n^2 + 2  n^4)
\right] 
	, \\
	\beta^{(2)}_{\Lambda} & = 
	-\frac{m^2 f^2 }{4} (n^2 - 1) (n^2 - 4).
	\label{eq:rg_SUn_2l}
\end{align}

To compare our results with that of ref.~\cite{Hnatic:2020kyo}, one has to take into account that the latter correspond to $f=0$ and are written in terms of $g_i/(8\pi^2)$ with  $g_1 = (n \lambda_2)/6$ and $g_2 = \lambda_1/6$. 
We also use the expressions obtained by means of \texttt{ARGES} \cite{Steudtner:2021fzs} to cross-check $\gamma_\phi$ and the beta functions for $\lambda_1$,  $\lambda_2$, and $m^2$ up to 4 loops.

\subsection{Higher rank $O(n) \times O(n) \times O(n)$ tensor model\label{sec:tensor}}

To give an example how to apply our  general result to models with
more complicated index structure, we consider evaluation of the beta functions
in the model with $O(n) \times O(n) \times O(n)$ symmetry
~\cite{Giombi:2017dtl}.
The model Lagrangian is
\begin{align}
  \label{eq:Lag_OnOnOn}
	\Lag =  \frac{1}{2} \partial_{\mu} \phi_{abc} \partial_{\mu} \phi_{abc}
  + \frac{\lambda_1}{4!} T^{\rm t}_{\phi_1 \phi_2 \phi_3 \phi_4}
  + \frac{\lambda_2}{4!} T^{\rm p}_{\phi_1 \phi_2 \phi_3 \phi_4}
  + \frac{\lambda_3}{4!} T^{\rm ds}_{\phi_1 \phi_2 \phi_3 \phi_4},
\end{align}
where we follow the naming scheme from ref.~\cite{Giombi:2017dtl} for the interaction terms as ``tetrahedral'',
``pillow'' and ``double-sum''. Again, we use fields as tensor indices of the structures $T^i$ to indicate contractions of triplets of indexes with $\phi_{abc}$.
For convenience, we present the tensor structures in the following pictorial form:
\tikzset{
  la/.style={line width=1pt,draw={rgb,255:red,224; green,66; blue,40}},
  lb/.style={line width=1pt,draw={rgb,255:red,120; green,145; blue,54}},
  lc/.style={line width=1pt,draw={rgb,255:red,21; green,68; blue,148}}
}
\begin{align}
  T^{\rm t}
  & = 
  \vcenter{\hbox{
      \begin{tikzpicture}[use Hobby shortcut, scale=0.5]
        \coordinate (va1i) at (-1,0.3);
        \coordinate (va2i) at (-0.3,-1);
        \coordinate (va3i) at (1,-0.3);
        \coordinate (va4i) at (0.3,1);
        \coordinate (va1o) at (-2,0.3);
        \coordinate (va2o) at (-0.3,-2);
        \coordinate (va3o) at (2,-0.3);
        \coordinate (va4o) at (0.3,2);
        \coordinate (vb1i) at (-1,0);
        \coordinate (vb2i) at (0,-1);
        \coordinate (vb3i) at (1,0);
        \coordinate (vb4i) at (0,1);
        \coordinate (vb1o) at (-2,0);
        \coordinate (vb2o) at (0,-2);
        \coordinate (vb3o) at (2,0);
        \coordinate (vb4o) at (0,2);
        \coordinate (vc1i) at (-1,-0.3);
        \coordinate (vc2i) at (0.3,-1);
        \coordinate (vc3i) at (1,0.3);
        \coordinate (vc4i) at (-0.3,1);
        \coordinate (vc1o) at (-2,-0.3);
        \coordinate (vc2o) at (0.3,-2);
        \coordinate (vc3o) at (2,0.3);
        \coordinate (vc4o) at (-0.3,2);
        \draw[la] (va1o) -- (va1i);
        \draw[la] (va2o) -- (va2i);
        \draw[la] (va3o) -- (va3i);
        \draw[la] (va4o) -- (va4i);
        \draw[lb] (vb1o) -- (vb1i);
        \draw[lb] (vb2o) -- (vb2i);
        \draw[lb] (vb3o) -- (vb3i);
        \draw[lb] (vb4o) -- (vb4i);
        \draw[lc] (vc1o) -- (vc1i);
        \draw[lc] (vc2o) -- (vc2i);
        \draw[lc] (vc3o) -- (vc3i);
        \draw[lc] (vc4o) -- (vc4i);
        \draw[la] (va1i) -- (va2i);
        \draw[la] (va3i) -- (va4i);
        \draw[lb] (vb1i) -- (vb3i);
        \draw[lb] (vb2i) -- (vb4i);
        \draw[lc] (vc1i) -- (vc4i);
        \draw[lc] (vc2i) -- (vc3i);
        \node[anchor=east] at (-2,0) {\tiny $\begin{matrix}
            a_1 \\
            b_1\\
            c_1
          \end{matrix}$};
        \node[anchor=north] at (0,-2) {\tiny $\begin{matrix}
            a_2 & b_2 & c_2
          \end{matrix}$};
        \node[anchor=west] at (2,0) {\tiny $\begin{matrix}
            c_3 \\
            b_3\\
            a_3
          \end{matrix}$};
        \node[anchor=south] at (0,2) {\tiny $\begin{matrix}
            c_4 & b_4 & a_4
          \end{matrix}$};
      \end{tikzpicture}
                        }},
                        \quad \quad
                        T^{\rm ds}
                        = 
                        \vcenter{\hbox{
      \begin{tikzpicture}[use Hobby shortcut, scale=0.5]
        \coordinate (va1i) at (-1,0.3);
        \coordinate (va2i) at (-0.3,-1);
        \coordinate (va3i) at (1,-0.3);
        \coordinate (va4i) at (0.3,1);
        \coordinate (va1o) at (-2,0.3);
        \coordinate (va2o) at (-0.3,-2);
        \coordinate (va3o) at (2,-0.3);
        \coordinate (va4o) at (0.3,2);
        \coordinate (vb1i) at (-1,0);
        \coordinate (vb2i) at (0,-1);
        \coordinate (vb3i) at (1,0);
        \coordinate (vb4i) at (0,1);
        \coordinate (vb1o) at (-2,0);
        \coordinate (vb2o) at (0,-2);
        \coordinate (vb3o) at (2,0);
        \coordinate (vb4o) at (0,2);
        \coordinate (vc1i) at (-1,-0.3);
        \coordinate (vc2i) at (0.3,-1);
        \coordinate (vc3i) at (1,0.3);
        \coordinate (vc4i) at (-0.3,1);
        \coordinate (vc1o) at (-2,-0.3);
        \coordinate (vc2o) at (0.3,-2);
        \coordinate (vc3o) at (2,0.3);
        \coordinate (vc4o) at (-0.3,2);
        \draw[la] (va1o) -- (va1i);
        \draw[la] (va2o) -- (va2i);
        \draw[la] (va3o) -- (va3i);
        \draw[la] (va4o) -- (va4i);
        \draw[lb] (vb1o) -- (vb1i);
        \draw[lb] (vb2o) -- (vb2i);
        \draw[lb] (vb3o) -- (vb3i);
        \draw[lb] (vb4o) -- (vb4i);
        \draw[lc] (vc1o) -- (vc1i);
        \draw[lc] (vc2o) -- (vc2i);
        \draw[lc] (vc3o) -- (vc3i);
        \draw[lc] (vc4o) -- (vc4i);
        \draw[la] (va1i) -- (va2i);
        \draw[la] (va3i) -- (va4i);
        \draw[lb] (vb1i) -- (vb2i);
        \draw[lb] (vb3i) -- (vb4i);
        \draw[lc] (vc1i) -- (vc2i);
        \draw[lc] (vc3i) -- (vc4i);
        \node[anchor=east] at (-2,0) {\tiny $\begin{matrix}
            a_1 \\
            b_1\\
            c_1
          \end{matrix}$};
        \node[anchor=north] at (0,-2) {\tiny $\begin{matrix}
            a_2 & b_2 & c_2
          \end{matrix}$};
        \node[anchor=west] at (2,0) {\tiny $\begin{matrix}
            c_3 \\
            b_3\\
            a_3
          \end{matrix}$};
        \node[anchor=south] at (0,2) {\tiny $\begin{matrix}
            c_4 & b_4 & a_4
          \end{matrix}$};
      \end{tikzpicture}
    }},
  \\
  T^{\rm p}
  & = \frac{1}{3} \left(
  \vcenter{\hbox{
      \begin{tikzpicture}[use Hobby shortcut, scale=0.5]
        \coordinate (va1i) at (-1,0.3);
        \coordinate (va2i) at (-0.3,-1);
        \coordinate (va3i) at (1,-0.3);
        \coordinate (va4i) at (0.3,1);
        \coordinate (va1o) at (-2,0.3);
        \coordinate (va2o) at (-0.3,-2);
        \coordinate (va3o) at (2,-0.3);
        \coordinate (va4o) at (0.3,2);
        \coordinate (vb1i) at (-1,0);
        \coordinate (vb2i) at (0,-1);
        \coordinate (vb3i) at (1,0);
        \coordinate (vb4i) at (0,1);
        \coordinate (vb1o) at (-2,0);
        \coordinate (vb2o) at (0,-2);
        \coordinate (vb3o) at (2,0);
        \coordinate (vb4o) at (0,2);
        \coordinate (vc1i) at (-1,-0.3);
        \coordinate (vc2i) at (0.3,-1);
        \coordinate (vc3i) at (1,0.3);
        \coordinate (vc4i) at (-0.3,1);
        \coordinate (vc1o) at (-2,-0.3);
        \coordinate (vc2o) at (0.3,-2);
        \coordinate (vc3o) at (2,0.3);
        \coordinate (vc4o) at (-0.3,2);
        \draw[la] (va1o) -- (va1i);
        \draw[la] (va2o) -- (va2i);
        \draw[la] (va3o) -- (va3i);
        \draw[la] (va4o) -- (va4i);
        \draw[lb] (vb1o) -- (vb1i);
        \draw[lb] (vb2o) -- (vb2i);
        \draw[lb] (vb3o) -- (vb3i);
        \draw[lb] (vb4o) -- (vb4i);
        \draw[lc] (vc1o) -- (vc1i);
        \draw[lc] (vc2o) -- (vc2i);
        \draw[lc] (vc3o) -- (vc3i);
        \draw[lc] (vc4o) -- (vc4i);
        \draw[la] (va1i) -- (va2i);
        \draw[la] (va3i) -- (va4i);
        \draw[lb] (vb1i) -- (vb2i);
        \draw[lb] (vb3i) -- (vb4i);
        \draw[lc] (vc1i) -- (vc4i);
        \draw[lc] (vc2i) -- (vc3i);
        \node[anchor=east] at (-2,0) {\tiny $\begin{matrix}
            a_1 \\
            b_1\\
            c_1
          \end{matrix}$};
        \node[anchor=north] at (0,-2) {\tiny $\begin{matrix}
            a_2 & b_2 & c_2
          \end{matrix}$};
        \node[anchor=west] at (2,0) {\tiny $\begin{matrix}
            c_3 \\
            b_3\\
            a_3
          \end{matrix}$};
        \node[anchor=south] at (0,2) {\tiny $\begin{matrix}
            c_4 & b_4 & a_4
          \end{matrix}$};
      \end{tikzpicture}
    }}
  +
  \vcenter{\hbox{
      \begin{tikzpicture}[use Hobby shortcut, scale=0.5]
        \coordinate (va1i) at (-1,0.3);
        \coordinate (va2i) at (-0.3,-1);
        \coordinate (va3i) at (1,-0.3);
        \coordinate (va4i) at (0.3,1);
        \coordinate (va1o) at (-2,0.3);
        \coordinate (va2o) at (-0.3,-2);
        \coordinate (va3o) at (2,-0.3);
        \coordinate (va4o) at (0.3,2);
        \coordinate (vb1i) at (-1,0);
        \coordinate (vb2i) at (0,-1);
        \coordinate (vb3i) at (1,0);
        \coordinate (vb4i) at (0,1);
        \coordinate (vb1o) at (-2,0);
        \coordinate (vb2o) at (0,-2);
        \coordinate (vb3o) at (2,0);
        \coordinate (vb4o) at (0,2);
        \coordinate (vc1i) at (-1,-0.3);
        \coordinate (vc2i) at (0.3,-1);
        \coordinate (vc3i) at (1,0.3);
        \coordinate (vc4i) at (-0.3,1);
        \coordinate (vc1o) at (-2,-0.3);
        \coordinate (vc2o) at (0.3,-2);
        \coordinate (vc3o) at (2,0.3);
        \coordinate (vc4o) at (-0.3,2);
        \draw[la] (va1o) -- (va1i);
        \draw[la] (va2o) -- (va2i);
        \draw[la] (va3o) -- (va3i);
        \draw[la] (va4o) -- (va4i);
        \draw[lb] (vb1o) -- (vb1i);
        \draw[lb] (vb2o) -- (vb2i);
        \draw[lb] (vb3o) -- (vb3i);
        \draw[lb] (vb4o) -- (vb4i);
        \draw[lc] (vc1o) -- (vc1i);
        \draw[lc] (vc2o) -- (vc2i);
        \draw[lc] (vc3o) -- (vc3i);
        \draw[lc] (vc4o) -- (vc4i);
        \draw[la] (va1i) -- (va4i);
        \draw[la] (va2i) -- (va3i);
        \draw[lb] (vb1i) -- (vb2i);
        \draw[lb] (vb3i) -- (vb4i);
        \draw[lc] (vc1i) -- (vc2i);
        \draw[lc] (vc3i) -- (vc4i);
        \node[anchor=east] at (-2,0) {\tiny $\begin{matrix}
            a_1 \\
            b_1\\
            c_1
          \end{matrix}$};
        \node[anchor=north] at (0,-2) {\tiny $\begin{matrix}
            a_2 & b_2 & c_2
          \end{matrix}$};
        \node[anchor=west] at (2,0) {\tiny $\begin{matrix}
            c_3 \\
            b_3\\
            a_3
          \end{matrix}$};
        \node[anchor=south] at (0,2) {\tiny $\begin{matrix}
            c_4 & b_4 & a_4
          \end{matrix}$};
      \end{tikzpicture}
    }}
  +
    \vcenter{\hbox{
      \begin{tikzpicture}[use Hobby shortcut, scale=0.5]
        \coordinate (va1i) at (-1,0.3);
        \coordinate (va2i) at (-0.3,-1);
        \coordinate (va3i) at (1,-0.3);
        \coordinate (va4i) at (0.3,1);
        \coordinate (va1o) at (-2,0.3);
        \coordinate (va2o) at (-0.3,-2);
        \coordinate (va3o) at (2,-0.3);
        \coordinate (va4o) at (0.3,2);
        \coordinate (vb1i) at (-1,0);
        \coordinate (vb2i) at (0,-1);
        \coordinate (vb3i) at (1,0);
        \coordinate (vb4i) at (0,1);
        \coordinate (vb1o) at (-2,0);
        \coordinate (vb2o) at (0,-2);
        \coordinate (vb3o) at (2,0);
        \coordinate (vb4o) at (0,2);
        \coordinate (vc1i) at (-1,-0.3);
        \coordinate (vc2i) at (0.3,-1);
        \coordinate (vc3i) at (1,0.3);
        \coordinate (vc4i) at (-0.3,1);
        \coordinate (vc1o) at (-2,-0.3);
        \coordinate (vc2o) at (0.3,-2);
        \coordinate (vc3o) at (2,0.3);
        \coordinate (vc4o) at (-0.3,2);
        \draw[la] (va1o) -- (va1i);
        \draw[la] (va2o) -- (va2i);
        \draw[la] (va3o) -- (va3i);
        \draw[la] (va4o) -- (va4i);
        \draw[lb] (vb1o) -- (vb1i);
        \draw[lb] (vb2o) -- (vb2i);
        \draw[lb] (vb3o) -- (vb3i);
        \draw[lb] (vb4o) -- (vb4i);
        \draw[lc] (vc1o) -- (vc1i);
        \draw[lc] (vc2o) -- (vc2i);
        \draw[lc] (vc3o) -- (vc3i);
        \draw[lc] (vc4o) -- (vc4i);
        \draw[la] (va1i) -- (va2i);
        \draw[la] (va3i) -- (va4i);
        \draw[lb] (vb1i) -- (vb3i);
        \draw[lb] (vb2i) -- (vb4i);
        \draw[lc] (vc1i) -- (vc2i);
        \draw[lc] (vc3i) -- (vc4i);
        \node[anchor=east] at (-2,0) {\tiny $\begin{matrix}
            a_1 \\
            b_1\\
            c_1
          \end{matrix}$};
        \node[anchor=north] at (0,-2) {\tiny $\begin{matrix}
            a_2 & b_2 & c_2
          \end{matrix}$};
        \node[anchor=west] at (2,0) {\tiny $\begin{matrix}
            c_3 \\
            b_3\\
            a_3
          \end{matrix}$};
        \node[anchor=south] at (0,2) {\tiny $\begin{matrix}
            c_4 & b_4 & a_4
          \end{matrix}$};
      \end{tikzpicture}
    }}
\right)
\end{align}
To map the model onto our general result, 
we associate open indices in general model \eqref{eq:Lag_general} with multi index $i_k=\{a_k,b_k,c_k\}$, and rewrite the self-coupling \eqref{eq:Lag_OnOnOn} in the form:
\begin{equation}
  \label{eq:lamOnOnOn}
  \lambda_{i_1i_2i_3i_4} = \lambda_1 T^{\rm t}_{(i_1 i_2 i_3 i_4)} + \lambda_2 T^{\rm p}_{(i_1 i_2 i_3 i_4)} + \lambda_3 T^{\rm ds}_{(i_1 i_2 i_3 i_4)}
\end{equation}
where $(i_1i_2i_3i_4)$ denotes symmetrization. 
At one loop we get
\begin{align}
	\beta^{(1)}_{1} & = 
	\frac{2 \lambda_1 \lambda_2}{3} (1 + n)
	+ 4 \lambda_1 \lambda_3 
	+ \frac{4 \lambda_2^2}{9},\\
	\beta^{(1)}_{2} & = 
	\lambda_1^2 (2+ n)
	+ \frac{4 \lambda_1 \lambda_2}{3} (2 + n)
	+ 4 \lambda_2 \lambda_3 
	+ \frac{\lambda_2^2}{9}(12 + 5 n + n^2), \\
	\beta^{(1)}_{3} & = 
	\frac{2 \lambda_1 \lambda_2}{3}
	+ 2 \lambda_1 \lambda_3 n  
	+ \frac{ 2 \lambda_2 \lambda_3 }{3} (1 + n + n^2)
	+ \frac{\lambda_2^2}{9} (3 + 2 n)
	+ \frac{\lambda_3^2}{3} (8 + n^3).
	\label{eq:rg_OnOnOn_1l}
\end{align}
while two-loop contribution renders
{\allowdisplaybreaks
\begin{align}
	\beta^{(2)}_{1} & = 
\frac{\lambda_1^3}{18} (n^3-15 n-10)
-\frac{2\lambda_1^2 \lambda_2}{9} (n^2+4 n+13)
-\frac{10\lambda_1^2 \lambda_3}{3} n
\nonumber\\
			&
-\frac{ \lambda_1 \lambda_2^2}{54} (n^3+15 n^2+93 n+101)
-\frac{2\lambda_1 \lambda_2\lambda_3 }{9}
(5 n^2+17 n+17)
\nonumber\\
			&
-\frac{\lambda_1 \lambda_3^2}{9}  (5 n^3+82)
-\frac{2\lambda_2^3}{81}  (2 n^2+13 n+24)
-\frac{16 \lambda_2^2 \lambda_3}{9},
\\
	\beta^{(2)}_{2} & = 
-\frac{2\lambda_1^3}{3}  (n^2+n+4)
-\frac{\lambda_1^2 \lambda_2}{18}  (n^3+12 n^2+99 n+98)
-4 \lambda_1^2 \lambda_3 (n+2)
\nonumber\\
			& -\frac{2\lambda_1 \lambda_2^2 }{9} (4 n^2+18 n+29)
-\frac{2\lambda_1 \lambda_2 \lambda_3}{3}  (13 n+16)
-\frac{\lambda_2^3 }{162} (5 n^3+45 n^2+243 n+343)
\nonumber\\
			&
-\frac{2\lambda_2^2 \lambda_3}{9}  (7 n^2+15 n+29)
-\frac{\lambda_2 \lambda_3^2 }{9} (5 n^3+82),
\\
	\beta^{(2)}_{3} & = 
-\frac{\lambda_1^3 n}{3}
-\frac{2\lambda_1^2 \lambda_2 }{9} (n^2+n+4)
-\frac{5\lambda_1^2 \lambda_3 }{18} (n^3+3 n+2)
-\frac{8\lambda_1 \lambda_2^2}{9}  (n+1)
\nonumber\\
			&
-\frac{2\lambda_1 \lambda_2 \lambda_3}{9}  (5 n^2+5 n+17)
-\frac{22\lambda_1 \lambda_3^2 }{3} n
-\frac{7\lambda_2^3 }{81} (n^2+3 n+5)
\nonumber\\
			& -\frac{\lambda_2^2 \lambda_3 }{54} (5 n^3+15 n^2+93 n+97)
-\frac{22\lambda_2 \lambda_3^2 }{9} (n^2+n+1)
-\frac{\lambda_3^3}{3}  (3 n^3+14).
	\label{eq:rg_OnOnOn_2l}
\end{align}
}

Modulo rescaling $\lambda_i = 6 g_i$, the obtained expressions coincide with those given in ref.~\cite{Giombi:2017dtl}. 
Six-loop\footnote{Due to the large number of indices in $\lambda_{i_1i_2i_3i_4}$ \eqref{eq:lamOnOnOn}, we use the JINR supercomputer ``Govorun'' to compute the required 571 six-loop tensors in parallel.} results can be found in the form of supplementary material. 

\subsection{Two-Higgs Doublet Model\label{sec:thdm}}

Motivated by three-loop calculation \cite{Bednyakov:2018cmx}  in 2HDM model (see, e.g., refs.~\cite{Branco:2011iw,Ivanov:2017dad} for review),
we consider the following general renormalizable Higgs potential 
\begin{align}
	V_{\mathrm{2HDM}} & = 
		  m_{11}^2 \Phi_1^\dagger \Phi_1
		+ m_{22}^2 \Phi_2^\dagger \Phi_2
		- \left(m_{12}^2 \Phi_1^\dagger \Phi_2 + \hc\right)
		\nonumber\\
		& 
		+ \frac{1}{2} \lambda_1 \left(\Phi_1^\dagger \Phi_1\right)^2 
		+ \frac{1}{2} \lambda_2 \left(\Phi_2^\dagger \Phi_2\right)^2 
		+  \lambda_3 \left(\Phi_1^\dagger \Phi_1\right)\left(\Phi_2^\dagger \Phi_2\right) 
		+  \lambda_4 \left(\Phi_1^\dagger \Phi_2\right)\left(\Phi_2^\dagger \Phi_1\right) 
		\nonumber\\
		&+ \left[ 
			\frac{1}{2} \lambda_5 \left(\Phi_1^\dagger \Phi_2\right)^2
			+ \lambda_6 \left(\Phi_1^\dagger \Phi_1\right) \left(\Phi_1^\dagger \Phi_2\right)
			+ \lambda_7 \left(\Phi_2^\dagger \Phi_2\right) \left(\Phi_1^\dagger \Phi_2\right)
			+ \hc
		\right],
\label{eq:thdm_potential}
\end{align}
where $\Phi_{1,2}$ are $SU(2)$ doublets. The self-couplings $\lambda_{1-4}$ and the mass parameters  $m_{11}^2$, $m_{22}^2$ are real, while $\lambda_{5-7}$, and $m_{12}^2$ can be complex. 
Due to the freedom in redefinition of Higgs-field basis,  only 11 of 14 real parameters in eq.~\eqref{eq:thdm_potential} are independent.
In ref.~\cite{Bednyakov:2018cmx} convenient variables \cite{Branco:2011iw,Ivanov:2017dad} and the so-called \emph{reparametrization invariants} (see, e.g., ref.~\cite{Trautner:2018ipq} for a comprehensive study) were used to compute the RG functions. 

In this work, we use another strategy and directly calculate the beta function of $\lambda_{1-7}$ together with the anomalous dimensions of $m^2_{11}$, $m_{22}^2$, and $m^2_{12}$ from our general expressions. 
We enumerate all real components of two doublets $\Phi_{1,2}$ and rewrite eq.~\eqref{eq:thdm_potential} in the general form \eqref{eq:Lag_general} with indices $a,b$, etc. running from one to eight. 
We find full agreement with previous results and extend the latter up to six loops. We have checked that our expressions for $\beta_{\lambda_2}(\beta_{\lambda_7})$ can be obtained from $\beta_{\lambda_1}(\beta_{\lambda_6})$ via the replacement $\lambda_1 \leftrightarrow \lambda_2$ and $\lambda_6 \leftrightarrow \lambda_7$.
One can use the same substitutions together with $m^2_{11} \leftrightarrow m^2_{22}$ to get $\beta_{m_{22}^2}$ from $\beta_{m_{11}^2}$.
We make the six-loop results available as supplementary material.

\section{Conclusion\label{sec:conclusion}}
We considered the general renormalizable scalar QFT model and directly computed
the RG functions for the quartic and cubic self-couplings, mass parameter,
tadpole term, and vacuum energy. In deriving our results for dimensionless
quantities, we used the expressions for the $\KRp$ operation applied to individual
Feynman integrals. The latter are publicly available thanks to lengthy and
non-trivial calculations of ref.~\cite{Kompaniets:2017yct}. To compute the RG functions of
dimensionful parameters, we utilize the powerful dummy field technique.

To validate our general results, we considered several scalar models discussed
in the theory of critical phenomena. We found perfect agreement with known
results and extend them by computing several missing six-loop contributions.
Among the latter are the vacuum energy beta function in the $O(n)$ model, the
anomalous dimensions of quadratic perturbations in the $O(n) \times O(m)$ model,
and the self-coupling beta functions for $U(n) \times U(m)$, and $O(n) \times
O(n) \times O(n)$ models and the model with the Higgs field in the adjoint
representation of the $SU(n)$ group. Additionally, we extend the three-loop
results for the general Two-Higgs-Doublet Model scalar sector to six loops.

We believe that the obtained state-of-the-art RG functions are of immediate
interest to the condensed-matter community. On the contrary, present six-loop
results can hardly find their applications in phenomenological analyses of the
Standard Model extensions in the near future. However, it is convenient to
estimate the influence of the high-order terms on extended Higgs sector studies,
which currently rely on the two- or three-loop RG. 
Public codes for RG
analyses~\cite{Staub:2013tta,Sartore:2020gou,Litim:2020jvl,Deppisch:2020aoj,Thomsen:2021ncy}
can be equipped with our results to carry out this kind of computations.

We also note that the expression for vacuum energy beta function is relevant for
effective potential $V_{\mathrm{eff}}(\phi)$ RG improvement~(see, e.g.,
ref.~\cite{Martin:2017lqn}). Moreover, in recent ref.~\cite{Manohar:2020nzp},
the vacuum energy function's role is emphasized in the effective field theory
approach to $V_{\mathrm{eff}}(\phi)$ computation in models with many different
scales.

Let us also mention here that seven-loop results~\cite{Schnetz:2016fhy} can not be directly used in our approach. We expect that in the future when the corresponding diagram-by-diagram counter-terms will be available, one can almost immediately extend our general expressions to one more loop. However, our experience tells us that the calculation of tensor structures in specific models can be very time-consuming.

\acknowledgments

We thank G.Kalagov, M.Kompaniets, N.Lebedev, and F. Herren for fruitful discussions. We also
thank T. Steudtner for the correspondence regarding
refs.~\cite{Steudtner:2020tzo, Steudtner:2021fzs} and sharing his four-loop
results. We are grateful to the Joint Institute for Nuclear Research for letting
us use their supercomputer ``Govorun''. The work of A.B. is supported by the
Grant of the Russian Federation Government, Agreement No. 14.W03.31.0026 from
15.02.2018. The work of A.P. is supported by the Foundation for the Advancement
of Theoretical Physics and Mathematics ``BASIS.''

\appendix
\pgfdeclarelayer{edgelayer}
\pgfdeclarelayer{nodelayer}
\pgfsetlayers{edgelayer,nodelayer,main}
\tikzstyle{none}=[inner sep=0pt]
\tikzstyle{white_circle}=[fill=white, draw=black, shape=circle, minimum size=14pt, inner sep=0pt]
\tikzstyle{white}=[fill=white, shape=circle, minimum size=12pt, inner sep=0pt]
\tikzstyle{lam}=[fill=cyan, draw=black, shape=circle, scale=0.8]
\tikzstyle{tri}=[fill=red, draw=black, shape=circle, scale=0.8]
\tikzstyle{mm}=[fill=green, draw=black, shape=circle, scale=0.8]
\tikzstyle{tad}=[fill=orange, draw=black, shape=circle, scale=0.8]
\tikzstyle{dummy}=[fill={rgb,255: red,191; green,128; blue,64}, draw=black, shape=circle, minimum size=14pt, inner sep=0pt]
\tikzstyle{dummy edge}=[dashed]
\tikzset{every picture/.style={baseline=-3pt,scale=0.7}}

\section{Deriving dimensionful couplings RG with Dummy field method\label{sec:dummy_field_method}}

\begin{figure}[t]
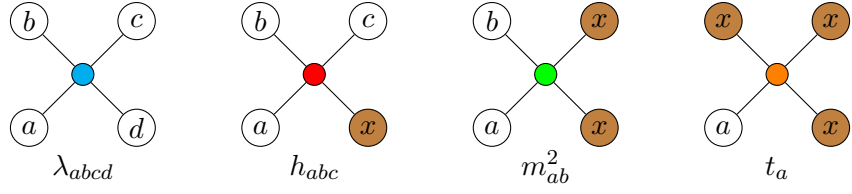

\begin{center}
\begin{tabular}{cccc}
	\tikz{
		\node[style=lam] (l) {};
		\node[style=white_circle, below left of=l] (a) {$a$}; 
		\node[style=white_circle, above left of=l] (b) {$b$}; 
		\node[style=white_circle, above right of=l] (c) {$c$}; 
		\node[style=white_circle, below right of=l] (d) {$d$}; 
		\draw (l) to (a);
		\draw (l) to (b);
		\draw (l) to (c);
		\draw (l) to (d);
	}
	&
	\tikz{
		\node[style=tri] (l) {};
		\node[style=white_circle, below left of=l] (a) {$a$}; 
		\node[style=white_circle, above left of=l] (b) {$b$}; 
		\node[style=white_circle, above right of=l] (c) {$c$}; 
		\node[style=dummy, below right of=l] (d) {$x$}; 
		\draw (l) to (a);
		\draw (l) to (b);
		\draw (l) to (c);
		\draw (l) to (d);
	}
	&
	\tikz{
		\node[style=mm] (l) {};
		\node[style=white_circle, below left of=l] (a) {$a$}; 
		\node[style=white_circle, above left of=l] (b) {$b$}; 
		\node[style=dummy, above right of=l] (c) {$x$}; 
		\node[style=dummy, below right of=l] (d) {$x$}; 
		\draw (l) to (a);
		\draw (l) to (b);
		\draw (l) to (c);
		\draw (l) to (d);
	}
	&
	\tikz{
		\node[style=tad] (l) {};
		\node[style=white_circle, below left of=l] (a) {$a$}; 
		\node[style=dummy, above left of=l] (b) {$x$}; 
		\node[style=dummy, above right of=l] (c) {$x$}; 
		\node[style=dummy, below right of=l] (d) {$x$}; 
		\draw (l) to (a);
		\draw (l) to (b);
		\draw (l) to (c);
		\draw (l) to (d);
	}
	\\
		$\lambda_{abcd}$
	& 
		$h_{abc}$ 
	& 
		$m^2_{ab}$ 
	&
		$t_{a}$
\end{tabular}
\end{center}
\caption{Representation of vertices corresponding to the parameters of the Lagrangian \eqref{eq:Lag_general}. 
The index ``$x$'' denotes the contraction with a dummy field.\label{tab:vertices}}
\end{figure}

The author of ref.~\cite{Steudtner:2020tzo} introduced a convenient representation for four-loop quartic-coupling beta function with all the self-couplings involving external indices explicitly factorized. 
 We adopt this ansatz to all loops
\begin{align}
	\beta_{abcd} & = 
		\left[
		  \lambda_{abcf} \gamma^\phi_{fd} 
		+ \lambda_{abdf} \gamma^\phi_{fc} 
		+ \lambda_{acdf} \gamma^\phi_{fb} 
		+ \lambda_{bcdf} \gamma^\phi_{fa} 
	\right] 
		     + 
		     \left[
			     \lambda_{abef} \lambda_{cdgh}
			     \bigcirc_{ef|gh}
			     + 5~\text{perm.}
		     \right]
	\nonumber\\
		     & +
		     \left[
			     \lambda_{abef} \lambda_{cghi} \lambda_{djkl}
		     \bigtriangleup_{ef|ghi|jkl}
			     + 11~\text{perm.}
		     \right] 
	\nonumber\\
		     & + 
		     \left[
			     \lambda_{aefg} \lambda_{bhij} \lambda_{cklm} \lambda_{dnop}
		     \square_{efg|hij|klm|nop}
			     + 23~\text{perm.}
		     \right]
\label{eq:beta_lambda_ctb}
\end{align}
and  re-derive the RG functions for the dimensionful parameters entering the Lagrangian~\eqref{eq:Lag_general}.
In eq.~\eqref{eq:beta_lambda_ctb} ``perm.'' denotes the terms, which can be obtained from the respective expressions via non-equivalent permutations of external indices. It is convenient to represent eq.~\eqref{eq:beta_lambda_ctb} in the following pictorial form: 
\begin{align}
	\beta_{abcd} & 
	=
	\left[
\begin{tikzpicture}
	\begin{pgfonlayer}{nodelayer}
		\node [style={white_circle}] (0) at (-2.5, 1.5) {$c$};
		\node [style={white_circle}] (1) at (-2.5, -1.5) {$a$};
		\node [style={white_circle}] (2) at (-2.5, 0) {$b$};
		\node [style={white_circle}] (3) at (2, 0) {$d$};
		\node [style=none] (4) at (-0.75, 0) {};
		\node [style=none] (5) at (0.75, 0) {};
		\node [style=lam] (9) at (-1.5, 0) {};
		\node [style=none] (10) at (0, 0) {};
		\node [style=none] (11) at (0, 0) {$\gamma$};
	\end{pgfonlayer}
	\begin{pgfonlayer}{edgelayer}
		\draw [bend left=90, looseness=1.50] (4.center) to (5.center);
		\draw [bend right=90, looseness=1.50] (4.center) to (5.center);
		\draw (0) to (9);
		\draw (9) to (1);
		\draw (9) to (2);
		\draw (9) to (4.center);
		\draw (5.center) to (3);
	\end{pgfonlayer}
\end{tikzpicture}
+ 3~\text{perm.}
	\right]
	+  
\left[
\begin{tikzpicture}
	\begin{pgfonlayer}{nodelayer}
		\node [style={white_circle}] (0) at (-2.25, 1.5) {$b$};
		\node [style={white_circle}] (1) at (-2.25, -1.5) {$a$};
		\node [style={white_circle}] (2) at (2.25, 1.5) {$c$};
		\node [style={white_circle}] (3) at (2.25, -1.5) {$d$};
		\node [style=none] (4) at (-0.75, 0) {};
		\node [style=none] (5) at (0.75, 0) {};
		\node [style=none] (6) at (-0.5, 0) {\scriptsize 1};
		\node [style=none] (7) at (0.5, 0) {\scriptsize 2};
		\node [style=lam] (18) at (1.25, 0) {};
		\node [style=lam] (21) at (-1.25, 0) {};
		\node [style=none] (57) at (-0.75, 0.25) {};
		\node [style=none] (58) at (-0.75, -0.25) {};
		\node [style=none] (59) at (0.75, -0.25) {};
		\node [style=none] (60) at (0.75, 0.25) {};
	\end{pgfonlayer}
	\begin{pgfonlayer}{edgelayer}
		\draw [bend left=90, looseness=1.50] (4.center) to (5.center);
		\draw [bend right=90, looseness=1.75] (4.center) to (5.center);
		\draw (2) to (18);
		\draw (18) to (3);
		\draw (0) to (21);
		\draw (21) to (1);
		\draw [bend left, looseness=0.75] (21) to (57.center);
		\draw [bend right] (21) to (58.center);
		\draw [bend right] (18) to (60.center);
		\draw [bend left] (18) to (59.center);
	\end{pgfonlayer}
\end{tikzpicture}
+ 5~\text{perm.}
\right]
\nonumber\\
		     & 
+
	\left[
\begin{tikzpicture}
	\begin{pgfonlayer}{nodelayer}
		\node [style={white_circle}] (22) at (-2, 1.5) {$b$};
		\node [style={white_circle}] (23) at (-2, -1.5) {$a$};
		\node [style={white_circle}] (24) at (1.75, 1.5) {$c$};
		\node [style={white_circle}] (25) at (1.75, -1.5) {$d$};
		\node [style=none] (26) at (-0.5, 0) {};
		\node [style=none] (28) at (0, 0) {\scriptsize 1};
		\node [style=none] (29) at (0.75, 0.25) {\scriptsize 2};
		\node [style=lam] (30) at (1.25, -1) {};
		\node [style=lam] (31) at (-0.75, 0) {};
		\node [style=none] (32) at (1, 0.75) {};
		\node [style=none] (33) at (1, -0.75) {};
		\node [style=lam] (34) at (1.25, 1) {};
		\node [style=none] (36) at (0.75, -0.25) {\scriptsize 3};
		\node [style=none] (61) at (1, 0.25) {};
		\node [style=none] (62) at (1, -0.25) {};
		\node [style=none] (63) at (0.5, -0.5) {};
		\node [style=none] (64) at (0.5, 0.5) {};
		\node [style=none] (65) at (0, 0.25) {};
		\node [style=none] (66) at (0, -0.25) {};
	\end{pgfonlayer}
	\begin{pgfonlayer}{edgelayer}
		\draw (22) to (31);
		\draw (31) to (23);
		\draw (26.center) to (32.center);
		\draw (32.center) to (33.center);
		\draw (33.center) to (26.center);
		\draw (32.center) to (34);
		\draw (34) to (24);
		\draw (33.center) to (30);
		\draw (30) to (25);
		\draw [bend left, looseness=1.25] (31) to (65.center);
		\draw [bend right, looseness=1.25] (31) to (66.center);
		\draw [bend left] (34) to (61.center);
		\draw [bend right] (34) to (64.center);
		\draw [bend right] (30) to (62.center);
		\draw [bend left] (30) to (63.center);
	\end{pgfonlayer}
\end{tikzpicture}
+ 11~\text{perm.}
\right] 
+
\left[
\begin{tikzpicture}
	\begin{pgfonlayer}{nodelayer}
		\node [style={white_circle}] (67) at (-2, 1.5) {$b$};
		\node [style={white_circle}] (68) at (-2, -1.5) {$a$};
		\node [style={white_circle}] (69) at (1, 1.5) {$c$};
		\node [style={white_circle}] (70) at (1, -1.5) {$d$};
		\node [style=none] (71) at (-1.25, -0.25) {};
		\node [style=none] (72) at (-1, -0.5) {\scriptsize 1};
		\node [style=none] (73) at (-1, 0.5) {\scriptsize 2};
		\node [style=lam] (74) at (0.5, -1) {};
		\node [style=lam] (75) at (-1.5, -1) {};
		\node [style=none] (76) at (0.25, 0.75) {};
		\node [style=none] (77) at (0.25, -0.75) {};
		\node [style=lam] (78) at (0.5, 1) {};
		\node [style=none] (79) at (0, 0.5) {\scriptsize 3};
		\node [style=none] (80) at (0.25, 0.25) {};
		\node [style=none] (81) at (0.25, -0.25) {};
		\node [style=none] (82) at (-0.25, -0.75) {};
		\node [style=none] (83) at (-0.25, 0.75) {};
		\node [style=none] (84) at (-1.25, 0.75) {};
		\node [style=none] (85) at (-1.25, -0.75) {};
		\node [style=none] (86) at (-0.75, -0.75) {};
		\node [style=none] (87) at (-1.25, 0.25) {};
		\node [style=none] (88) at (-0.75, 0.75) {};
		\node [style=lam] (89) at (-1.5, 1) {};
		\node [style=none] (90) at (0, -0.5) {};
		\node [style=none] (91) at (0, -0.5) {\scriptsize 4};
		\node [style=none] (92) at (0.25, 0.25) {};
	\end{pgfonlayer}
	\begin{pgfonlayer}{edgelayer}
		\draw (75) to (68);
		\draw (76.center) to (77.center);
		\draw (76.center) to (78);
		\draw (78) to (69);
		\draw (77.center) to (74);
		\draw (74) to (70);
		\draw [bend left] (78) to (80.center);
		\draw [bend right] (78) to (83.center);
		\draw [bend right] (74) to (81.center);
		\draw [bend left] (74) to (82.center);
		\draw (76.center) to (84.center);
		\draw (84.center) to (85.center);
		\draw (85.center) to (77.center);
		\draw [bend right, looseness=1.25] (75) to (86.center);
		\draw [bend left] (75) to (71.center);
		\draw (75) to (85.center);
		\draw [bend right, looseness=1.25] (89) to (87.center);
		\draw [bend left] (89) to (88.center);
		\draw (84.center) to (89);
		\draw (67) to (89);
	\end{pgfonlayer}
\end{tikzpicture}
+ 23~\text{perm.}
\right],
\label{eq:beta_lambda_ctb_graph}
\end{align}
where external self-couplings are denoted by blue vertices (see figure~\ref{tab:vertices}) and we 
use the notation given in figure~\ref{tab:ctb_notation} for the non-external parts of four-point functions. 
It is worth noting that $\bigcirc_{ab|cd}$,  $\bigtriangleup_{ab|cde|fgh}$, and $\square_{abc|def|ghi|jkl}$ do not need to be symmetric w.r.t. permutations of (group of) indices. 
Each group of indices is contracted with symmetric couplings, and, thus, 
does not need to be explicitly symmetrized. However, we explicitly take into account that an external index $a,b,c$, or $d$ can be attached to any group via a quartic vertex (this corresponds the permutations indicated, e.g., in eq.~\eqref{eq:beta_lambda_ctb_graph}).
Due to this, we distinguish index groups and mark them by numbers (c.f., figure~\ref{tab:ctb_notation}). 

\addtolength{\tabcolsep}{10pt}

\begin{figure}[t]
\begin{center}
\begin{tabular}{ccc}
\begin{tikzpicture}
	\begin{pgfonlayer}{nodelayer}
		\node [style=none] (4) at (-0.75, 0) {};
		\node [style=none] (5) at (0.75, 0) {};
		\node [style=none] (6) at (-0.5, 0) {\scriptsize 1};
		\node [style=none] (7) at (0.5, 0) {\scriptsize 2};
		\node [style=none] (57) at (-0.75, 0.25) {};
		\node [style=none] (58) at (-0.75, -0.25) {};
		\node [style=none] (59) at (0.75, -0.25) {};
		\node [style=none] (60) at (0.75, 0.25) {};
		\node [style=white] (61) at (-1.25, 0.75) {$e$};
		\node [style=white] (62) at (-1.25, -0.75) {$f$};
		\node [style=white] (63) at (1.25, 0.75) {$g$};
		\node [style=white] (64) at (1.25, -0.75) {$h$};
	\end{pgfonlayer}
	\begin{pgfonlayer}{edgelayer}
		\draw [bend left=90, looseness=1.50] (4.center) to (5.center);
		\draw [bend right=90, looseness=1.75] (4.center) to (5.center);
		\draw (57.center) to (61.center);
		\draw (58.center) to (62.center);
		\draw (60.center) to (63.center);
		\draw (59.center) to (64.center);
	\end{pgfonlayer}
\end{tikzpicture}
&
\begin{tikzpicture}
	\begin{pgfonlayer}{nodelayer}
		\node [style=none] (ef) at (-0.5, 0) {};
		\node [style=none] (28) at (0, 0) {\scriptsize 1};
		\node [style=none] (29) at (0.75, 0.32) {\scriptsize 2};
		\node [style=none] (ghi) at (1, 0.75) {};
		\node [style=none] (jkl) at (1, -0.75) {};
		\node [style=none] (36) at (0.75, -0.32) {\scriptsize 3};

		\node [style=none] (61) at ($(ghi.center)!0.10!(jkl.center)$) {};
		\node [style=none] (62) at ($(ghi.center)!0.90!(jkl.center)$) {};

		\node [style=none] (63) at ($(ef.center)!0.90!(jkl.center)$) {};

		\node [style=none] (64) at ($(ef.center)!0.90!(ghi.center)$) {};
		\node [style=none] (65) at ($(ef.center)!0.10!(ghi.center)$) {};
		\node [style=none] (66) at ($(ef.center)!0.10!(jkl.center)$) {};
		\node [style=white] (67) at (-0.75, 0.5) {$e$};
		\node [style=white] (68) at (-0.75, -0.5) {$f$};
		\node [style=white] (69) at (1.5, -1.25) {$k$};
		\node [style=white] (70) at (1.5, 1.25) {$h$};
		\node [style=white] (71) at (0.5, 1.25) {$g$};
		\node [style=white] (72) at (1.75, 0.5) {$i$};
		\node [style=white] (73) at (1.75, -0.5) {$l$};
		\node [style=white] (74) at (0.5, -1.25) {$j$};
	\end{pgfonlayer}
	\begin{pgfonlayer}{edgelayer}
		\draw (ef.center) to (ghi.center);
		\draw (ghi.center) to (jkl.center);
		\draw (jkl.center) to (ef.center);
		\draw (65.center) to (67.center);
		\draw (66.center) to (68.center);
		\draw (64.center) to (71.center);
		\draw (ghi.center) to (70.center);
		\draw (61.center) to (72.center);
		\draw (62.center) to (73.center);
		\draw (jkl.center) to (69.center);
		\draw (63.center) to (74.center);
	\end{pgfonlayer}
\end{tikzpicture}
&
\begin{tikzpicture}
	\begin{pgfonlayer}{nodelayer}
		\node [style=none] (72) at (-1, -0.5) {\scriptsize 1};
		\node [style=none] (73) at (-1, 0.5) {\scriptsize 2};
		\node [style=none] (klm) at (0.25, 0.75) {};
		\node [style=none] (nop) at (0.25, -0.75) {};
		\node [style=none] (79) at (0, 0.5) {\scriptsize 3};

		\node [style=none] (81) at ($(klm.center)!0.90!(nop.center)$) {};
		\node [style=none] (92) at ($(klm.center)!0.10!(nop.center)$) {};

		\node [style=none] (hij) at (-1.25, 0.75) {};
		\node [style=none] (efg) at (-1.25, -0.75) {};

		\node [style=none] (71) at ($(hij.center)!0.90!(efg.center)$) {};
		\node [style=none] (87) at ($(hij.center)!0.10!(efg.center)$) {};

		\node [style=none] (83) at ($(hij.center)!0.90!(klm.center)$) {};
		\node [style=none] (88) at ($(hij.center)!0.10!(klm.center)$) {};

		\node [style=none] (82) at ($(efg.center)!0.90!(nop.center)$) {};
		\node [style=none] (86) at ($(efg.center)!0.10!(nop.center)$) {};

		\node [style=none] (90) at (0, -0.5) {};
		\node [style=none] (91) at (0, -0.5) {\scriptsize 4};
		\node [style={white}] (93) at (-1.75, -0.5) {$g$};
		\node [style={white}] (94) at (-1.75, -1.25) {$f$};
		\node [style={white}] (95) at (-1, -1.25) {$e$};
		\node [style={white}] (96) at (0.75, -1.25) {$o$};
		\node [style={white}] (97) at (0, -1.25) {$p$};
		\node [style={white}] (98) at (0.75, -0.5) {$n$};
		\node [style={white}] (99) at (0.75, 0.5) {$m$};
		\node [style={white}] (100) at (0.75, 1.25) {$l$};
		\node [style={white}] (101) at (0, 1.25) {$k$};
		\node [style={white}] (102) at (-1, 1.25) {$j$};
		\node [style={white}] (103) at (-1.75, 1.25) {$i$};
		\node [style={white}] (104) at (-1.75, 0.5) {$h$};
	\end{pgfonlayer}
	\begin{pgfonlayer}{edgelayer}
		\draw (klm.center) to (nop.center);
		\draw (klm.center) to (hij.center);
		\draw (hij.center) to (efg.center);
		\draw (efg.center) to (nop.center);
		\draw (hij.center) to (103);
		\draw (87.center) to (104);
		\draw (88.center) to (102);
		\draw (83.center) to (101);
		\draw (klm.center) to (100);
		\draw (92.center) to (99);
		\draw (81.center) to (98);
		\draw (nop.center) to (96);
		\draw (82.center) to (97);
		\draw (86.center) to (95);
		\draw (efg.center) to (94);
		\draw (71.center) to (93);
	\end{pgfonlayer}
\end{tikzpicture}
\\
\end{tabular}
\addtolength{\tabcolsep}{-10pt}
\end{center}
\caption{Graphical representation of the 
structures entering  $\beta_{abcd}$ \eqref{eq:beta_lambda_ctb}.
The numbers encode the positions of index groups (separated by vertical lines) 
in the corresponding expressions.
\label{tab:ctb_notation}}
\end{figure}

Let us now contract the expression \eqref{eq:beta_lambda_ctb_graph} with external dummy field $x_d$
and exclude the tadpole graphs discussed in sec.~\ref{sec:details}: 
\begin{align}
	\beta_{abc} & 
	=
	\left[
%
\right)
+ 5~\text{perm.}
\right]
\label{eq:beta_tri_ctb_graph}
\end{align}
Here the trilinear couplings correspond to red vertices (see figure~\ref{tab:vertices}) and again we have to explicitly take into account 
permutations of external indices. The analytic expression is given by 
\begin{align}
	 \beta_{abc}  & = 
		\left[
		  h_{abf} \gamma^\phi_{fc} 
		  + h_{acf} \gamma^\phi_{fb} 
		  + h_{bcf} \gamma^\phi_{fa} 
	\right] 
		     + 
		     \left[
			     \lambda_{abef} h_{cgh}
			     \left(
				     \bigcirc_{ef|gh}
				     +
				     \bigcirc_{gh|ef}
		     	     \right)
			     + 2~\text{perm.}
		     \right]
	\nonumber\\
		     & +
		     \left[
			     \lambda_{abef} \lambda_{cghi} h_{jkl}
			     \left(
		     \bigtriangleup_{ef|ghi|jkl}
		     +
		     \bigtriangleup_{ef|jkl|ghi}
	     \right)
			     + 2~\text{perm.}
		     \right]
		     \nonumber\\
		     &
		     +
		     \left[
			     h_{aef} \lambda_{bghi} \lambda_{cjkl}
		     \bigtriangleup_{ef|ghi|jkl}
			     + 5~\text{perm.}
		     \right]
		     +
		     \left[
			     \lambda_{aefg} \lambda_{bhij} \lambda_{cklm} h_{nop}
			     \left(
		       \square_{efg|hij|klm|nop}
		     \right.
		     \right.
		     \nonumber\\
		     & 
		     \left. \left.
		     + \square_{nop|efg|hij|klm}
		     + \square_{klm|nop|efg|hij}
		     + \square_{hij|klm|nop|efg}
	     		     \right)
			     + 5~\text{perm.}
		     \right]
\label{eq:beta_tri_ctb}
\end{align}
	To obtain the beta function for mass parameter we contract eq.~\eqref{eq:beta_tri_ctb_graph} with one more dummy field $x_c$.
	Dividing the result by the factor of two, we get
{\allowdisplaybreaks
\begin{align}
	\beta_{ab} & 
	=
	\left[

\right) + (a\leftrightarrow b)
\right],
\label{eq:beta_mm_ctb_graph}
\end{align}
}
where  red dots denote mass parameter $m^2_{ab}$ insertions (c.f. figure~\ref{tab:vertices}).
The corresponding analytic expression is given by\footnote{We correct a couple of misprints in the corresponding expression in the published version of ref.~\cite{Steudtner:2020tzo}. }
\begin{align}
	\beta_{ab} & = 
		\left[
		  m^2_{af} \gamma^\phi_{fb} 
		  +
		  m^2_{bf} \gamma^\phi_{fa} 
	\right] 
		     + 
		     \left[
			     \left(\lambda_{abef} m^2_{gh} + h_{aef} h_{bgh}\right)
			     \left(
				     \bigcirc_{ef|gh}
				     +
				     \bigcirc_{gh|ef}
		     	     \right)
		     \right]
	\nonumber\\
		     & +
		     \left[
			     \lambda_{abef} h_{ghi} h_{jkl}
		     	      \bigtriangleup_{ef|ghi|jkl}
			      + 
			      m^2_{ef} \lambda_{aghi} \lambda_{bjkl}
			     \left(
				     \bigtriangleup_{ef|ghi|jkl}
				     +
				     \bigtriangleup_{ef|jkl|ghi}
	     		\right)
		     \right]
		     \nonumber\\
		     & 
		     \left[
			     h_{aef} h_{ghi} \lambda_{bjkl}
			     \left(
				     \bigtriangleup_{ef|ghi|jkl}
				     +
				     \bigtriangleup_{ef|jkl|ghi}
	     		\right)
			+ (a\leftrightarrow b)
		     \right]
		     \nonumber\\
		     &
		     +
		     \left[
			     \lambda_{aefg} \lambda_{bhij} h_{klm} h_{nop}
			     \left(
		         \square_{efg|hij|klm|nop} 
		       + \square_{efg|klm|hij|nop} 
		       + \square_{efg|klm|nop|hij} 
	       \right.
       \right.
       \nonumber\\
		     & 
		     \left. \left.
		       + \square_{klm|efg|hij|nop} 
		       + \square_{klm|efg|nop|hij} 
		       + \square_{klm|nop|efg|hij} 
	     		     \right)
			     + (a\leftrightarrow b)
		     \right].
		     \label{eq:beta_mm_ctb}
\end{align}

We proceed further and obtain the RG function for the tadpole term. 
Contracting eq.~\eqref{eq:beta_mm_ctb_graph} with $x_b$ and dividing by the factor of 3, we get
\begin{align}
	\beta_{a} & 
	=
	\left[
\begin{tikzpicture}
	\begin{pgfonlayer}{nodelayer}
		\node [style={dummy}] (0) at (-2.5, 1.5) {$x$};
		\node [style={dummy}] (1) at (-2.5, -1.5) {$x$};
		\node [style={dummy}] (2) at (-2.5, 0) {$x$};
		\node [style={white_circle}] (3) at (2, 0) {$a$};
		\node [style=none] (4) at (-0.75, 0) {};
		\node [style=none] (5) at (0.75, 0) {};
		\node [style=tad] (9) at (-1.5, 0) {};
		\node [style=none] (10) at (0, 0) {};
		\node [style=none] (11) at (0, 0) {$\gamma$};
	\end{pgfonlayer}
	\begin{pgfonlayer}{edgelayer}
		\draw [bend left=90, looseness=1.50] (4.center) to (5.center);
		\draw [bend right=90, looseness=1.50] (4.center) to (5.center);
		\draw (0) to (9);
		\draw (9) to (1);
		\draw (9) to (2);
		\draw (9) to (4.center);
		\draw (5.center) to (3);
	\end{pgfonlayer}
\end{tikzpicture}
\right]
+
\left[
\begin{tikzpicture}
	\begin{pgfonlayer}{nodelayer}
		\node [style={dummy}] (0) at (-2.25, 1.5) {$x$};
		\node [style={white_circle}] (1) at (-2.25, -1.5) {$a$};
		\node [style={dummy}] (2) at (2.25, 1.5) {$x$};
		\node [style={dummy}] (3) at (2.25, -1.5) {$x$};
		\node [style=none] (4) at (-0.75, 0) {};
		\node [style=none] (5) at (0.75, 0) {};
		\node [style=none] (6) at (-0.5, 0) {\scriptsize 1};
		\node [style=none] (7) at (0.5, 0) {\scriptsize 2};
		\node [style=mm] (18) at (1.25, 0) {};
		\node [style=tri] (21) at (-1.25, 0) {};
		\node [style=none] (57) at (-0.75, 0.25) {};
		\node [style=none] (58) at (-0.75, -0.25) {};
		\node [style=none] (59) at (0.75, -0.25) {};
		\node [style=none] (60) at (0.75, 0.25) {};
	\end{pgfonlayer}
	\begin{pgfonlayer}{edgelayer}
		\draw [bend left=90, looseness=1.50] (4.center) to (5.center);
		\draw [bend right=90, looseness=1.75] (4.center) to (5.center);
		\draw (2) to (18);
		\draw (18) to (3);
		\draw (0) to (21);
		\draw (21) to (1);
		\draw [bend left, looseness=0.75] (21) to (57.center);
		\draw [bend right] (21) to (58.center);
		\draw [bend right] (18) to (60.center);
		\draw [bend left] (18) to (59.center);
	\end{pgfonlayer}
\end{tikzpicture}
+
\begin{tikzpicture}
	\begin{pgfonlayer}{nodelayer}
		\node [style={dummy}] (0) at (-2.25, 1.5) {$x$};
		\node [style={dummy}] (1) at (-2.25, -1.5) {$x$};
		\node [style={dummy}] (2) at (2.25, 1.5) {$x$};
		\node [style={white_circle}] (3) at (2.25, -1.5) {$a$};
		\node [style=none] (4) at (-0.75, 0) {};
		\node [style=none] (5) at (0.75, 0) {};
		\node [style=none] (6) at (-0.5, 0) {\scriptsize 1};
		\node [style=none] (7) at (0.5, 0) {\scriptsize 2};
		\node [style=tri] (18) at (1.25, 0) {};
		\node [style=mm] (21) at (-1.25, 0) {};
		\node [style=none] (57) at (-0.75, 0.25) {};
		\node [style=none] (58) at (-0.75, -0.25) {};
		\node [style=none] (59) at (0.75, -0.25) {};
		\node [style=none] (60) at (0.75, 0.25) {};
	\end{pgfonlayer}
	\begin{pgfonlayer}{edgelayer}
		\draw [bend left=90, looseness=1.50] (4.center) to (5.center);
		\draw [bend right=90, looseness=1.75] (4.center) to (5.center);
		\draw (2) to (18);
		\draw (18) to (3);
		\draw (0) to (21);
		\draw (21) to (1);
		\draw [bend left, looseness=0.75] (21) to (57.center);
		\draw [bend right] (21) to (58.center);
		\draw [bend right] (18) to (60.center);
		\draw [bend left] (18) to (59.center);
	\end{pgfonlayer}
\end{tikzpicture}
\right]
\nonumber\\
		     & 
+
	\left[
\begin{tikzpicture}
	\begin{pgfonlayer}{nodelayer}
		\node [style={dummy}] (22) at (-2, 1.5) {$x$};
		\node [style={white_circle}] (23) at (-2, -1.5) {$a$};
		\node [style={dummy}] (24) at (1.75, 1.5) {$x$};
		\node [style={dummy}] (25) at (1.75, -1.5) {$x$};
		\node [style=none] (26) at (-0.5, 0) {};
		\node [style=none] (28) at (0, 0) {\scriptsize 1};
		\node [style=none] (29) at (0.75, 0.25) {\scriptsize 2};
		\node [style=tri] (30) at (1.25, -1) {};
		\node [style=tri] (31) at (-0.75, 0) {};
		\node [style=none] (32) at (1, 0.75) {};
		\node [style=none] (33) at (1, -0.75) {};
		\node [style=tri] (34) at (1.25, 1) {};
		\node [style=none] (36) at (0.75, -0.25) {\scriptsize 3};
		\node [style=none] (61) at (1, 0.25) {};
		\node [style=none] (62) at (1, -0.25) {};
		\node [style=none] (63) at (0.5, -0.5) {};
		\node [style=none] (64) at (0.5, 0.5) {};
		\node [style=none] (65) at (0, 0.25) {};
		\node [style=none] (66) at (0, -0.25) {};
	\end{pgfonlayer}
	\begin{pgfonlayer}{edgelayer}
		\draw (22) to (31);
		\draw (31) to (23);
		\draw (26.center) to (32.center);
		\draw (32.center) to (33.center);
		\draw (33.center) to (26.center);
		\draw (32.center) to (34);
		\draw (34) to (24);
		\draw (33.center) to (30);
		\draw (30) to (25);
		\draw [bend left, looseness=1.25] (31) to (65.center);
		\draw [bend right, looseness=1.25] (31) to (66.center);
		\draw [bend left] (34) to (61.center);
		\draw [bend right] (34) to (64.center);
		\draw [bend right] (30) to (62.center);
		\draw [bend left] (30) to (63.center);
	\end{pgfonlayer}
\end{tikzpicture}
+
\begin{tikzpicture}
	\begin{pgfonlayer}{nodelayer}
		\node [style={dummy}] (22) at (-2, 1.5) {$x$};
		\node [style={dummy}] (23) at (-2, -1.5) {$x$};
		\node [style={white_circle}] (24) at (1.75, 1.5) {$a$};
		\node [style={dummy}] (25) at (1.75, -1.5) {$x$};
		\node [style=none] (26) at (-0.5, 0) {};
		\node [style=none] (28) at (0, 0) {\scriptsize 1};
		\node [style=none] (29) at (0.75, 0.25) {\scriptsize 2};
		\node [style=tri] (30) at (1.25, -1) {};
		\node [style=mm] (31) at (-0.75, 0) {};
		\node [style=none] (32) at (1, 0.75) {};
		\node [style=none] (33) at (1, -0.75) {};
		\node [style=lam] (34) at (1.25, 1) {};
		\node [style=none] (36) at (0.75, -0.25) {\scriptsize 3};
		\node [style=none] (61) at (1, 0.25) {};
		\node [style=none] (62) at (1, -0.25) {};
		\node [style=none] (63) at (0.5, -0.5) {};
		\node [style=none] (64) at (0.5, 0.5) {};
		\node [style=none] (65) at (0, 0.25) {};
		\node [style=none] (66) at (0, -0.25) {};
	\end{pgfonlayer}
	\begin{pgfonlayer}{edgelayer}
		\draw (22) to (31);
		\draw (31) to (23);
		\draw (26.center) to (32.center);
		\draw (32.center) to (33.center);
		\draw (33.center) to (26.center);
		\draw (32.center) to (34);
		\draw (34) to (24);
		\draw (33.center) to (30);
		\draw (30) to (25);
		\draw [bend left, looseness=1.25] (31) to (65.center);
		\draw [bend right, looseness=1.25] (31) to (66.center);
		\draw [bend left] (34) to (61.center);
		\draw [bend right] (34) to (64.center);
		\draw [bend right] (30) to (62.center);
		\draw [bend left] (30) to (63.center);
	\end{pgfonlayer}
\end{tikzpicture}
+
\begin{tikzpicture}
	\begin{pgfonlayer}{nodelayer}
		\node [style={dummy}] (22) at (-2, 1.5) {$x$};
		\node [style={dummy}] (23) at (-2, -1.5) {$x$};
		\node [style={dummy}] (24) at (1.75, 1.5) {$x$};
		\node [style={white_circle}] (25) at (1.75, -1.5) {$a$};
		\node [style=none] (26) at (-0.5, 0) {};
		\node [style=none] (28) at (0, 0) {\scriptsize 1};
		\node [style=none] (29) at (0.75, 0.25) {\scriptsize 2};
		\node [style=lam] (30) at (1.25, -1) {};
		\node [style=mm] (31) at (-0.75, 0) {};
		\node [style=none] (32) at (1, 0.75) {};
		\node [style=none] (33) at (1, -0.75) {};
		\node [style=tri] (34) at (1.25, 1) {};
		\node [style=none] (36) at (0.75, -0.25) {\scriptsize 3};
		\node [style=none] (61) at (1, 0.25) {};
		\node [style=none] (62) at (1, -0.25) {};
		\node [style=none] (63) at (0.5, -0.5) {};
		\node [style=none] (64) at (0.5, 0.5) {};
		\node [style=none] (65) at (0, 0.25) {};
		\node [style=none] (66) at (0, -0.25) {};
	\end{pgfonlayer}
	\begin{pgfonlayer}{edgelayer}
		\draw (22) to (31);
		\draw (31) to (23);
		\draw (26.center) to (32.center);
		\draw (32.center) to (33.center);
		\draw (33.center) to (26.center);
		\draw (32.center) to (34);
		\draw (34) to (24);
		\draw (33.center) to (30);
		\draw (30) to (25);
		\draw [bend left, looseness=1.25] (31) to (65.center);
		\draw [bend right, looseness=1.25] (31) to (66.center);
		\draw [bend left] (34) to (61.center);
		\draw [bend right] (34) to (64.center);
		\draw [bend right] (30) to (62.center);
		\draw [bend left] (30) to (63.center);
	\end{pgfonlayer}
\end{tikzpicture}
\right] 
\nonumber\\
		     &+
\left[
	\left(
\begin{tikzpicture} 
	\begin{pgfonlayer}{nodelayer}
		\node [style={dummy}] (67) at (-2, 1.5) {$x$};
		\node [style={white_circle}] (68) at (-2, -1.5) {$a$};
		\node [style={dummy}] (69) at (1, 1.5) {$x$};
		\node [style={dummy}] (70) at (1, -1.5) {$x$};
		\node [style=none] (71) at (-1.25, -0.25) {};
		\node [style=none] (72) at (-1, -0.5) {\scriptsize 1};
		\node [style=none] (73) at (-1, 0.5) {\scriptsize 2};
		\node [style=tri] (74) at (0.5, -1) {};
		\node [style=lam] (75) at (-1.5, -1) {};
		\node [style=none] (76) at (0.25, 0.75) {};
		\node [style=none] (77) at (0.25, -0.75) {};
		\node [style=tri] (78) at (0.5, 1) {};
		\node [style=none] (79) at (0, 0.5) {\scriptsize 3};
		\node [style=none] (80) at (0.25, 0.25) {};
		\node [style=none] (81) at (0.25, -0.25) {};
		\node [style=none] (82) at (-0.25, -0.75) {};
		\node [style=none] (83) at (-0.25, 0.75) {};
		\node [style=none] (84) at (-1.25, 0.75) {};
		\node [style=none] (85) at (-1.25, -0.75) {};
		\node [style=none] (86) at (-0.75, -0.75) {};
		\node [style=none] (87) at (-1.25, 0.25) {};
		\node [style=none] (88) at (-0.75, 0.75) {};
		\node [style=tri] (89) at (-1.5, 1) {};
		\node [style=none] (90) at (0, -0.5) {};
		\node [style=none] (91) at (0, -0.5) {\scriptsize 4};
		\node [style=none] (92) at (0.25, 0.25) {};
	\end{pgfonlayer}
	\begin{pgfonlayer}{edgelayer}
		\draw (75) to (68);
		\draw (76.center) to (77.center);
		\draw (76.center) to (78);
		\draw (78) to (69);
		\draw (77.center) to (74);
		\draw (74) to (70);
		\draw [bend left] (78) to (80.center);
		\draw [bend right] (78) to (83.center);
		\draw [bend right] (74) to (81.center);
		\draw [bend left] (74) to (82.center);
		\draw (76.center) to (84.center);
		\draw (84.center) to (85.center);
		\draw (85.center) to (77.center);
		\draw [bend right, looseness=1.25] (75) to (86.center);
		\draw [bend left] (75) to (71.center);
		\draw (75) to (85.center);
		\draw [bend right, looseness=1.25] (89) to (87.center);
		\draw [bend left] (89) to (88.center);
		\draw (84.center) to (89);
		\draw (67) to (89);
	\end{pgfonlayer}
\end{tikzpicture}
+
\begin{tikzpicture} 
	\begin{pgfonlayer}{nodelayer}
		\node [style={white_circle}] (67) at (-2, 1.5) {$a$};
		\node [style={dummy}] (68) at (-2, -1.5) {$x$};
		\node [style={dummy}] (69) at (1, 1.5) {$x$};
		\node [style={dummy}] (70) at (1, -1.5) {$x$};
		\node [style=none] (71) at (-1.25, -0.25) {};
		\node [style=none] (72) at (-1, -0.5) {\scriptsize 1};
		\node [style=none] (73) at (-1, 0.5) {\scriptsize 2};
		\node [style=tri] (74) at (0.5, -1) {};
		\node [style=tri] (75) at (-1.5, -1) {};
		\node [style=none] (76) at (0.25, 0.75) {};
		\node [style=none] (77) at (0.25, -0.75) {};
		\node [style=tri] (78) at (0.5, 1) {};
		\node [style=none] (79) at (0, 0.5) {\scriptsize 3};
		\node [style=none] (80) at (0.25, 0.25) {};
		\node [style=none] (81) at (0.25, -0.25) {};
		\node [style=none] (82) at (-0.25, -0.75) {};
		\node [style=none] (83) at (-0.25, 0.75) {};
		\node [style=none] (84) at (-1.25, 0.75) {};
		\node [style=none] (85) at (-1.25, -0.75) {};
		\node [style=none] (86) at (-0.75, -0.75) {};
		\node [style=none] (87) at (-1.25, 0.25) {};
		\node [style=none] (88) at (-0.75, 0.75) {};
		\node [style=lam] (89) at (-1.5, 1) {};
		\node [style=none] (90) at (0, -0.5) {};
		\node [style=none] (91) at (0, -0.5) {\scriptsize 4};
		\node [style=none] (92) at (0.25, 0.25) {};
	\end{pgfonlayer}
	\begin{pgfonlayer}{edgelayer}
		\draw (75) to (68);
		\draw (76.center) to (77.center);
		\draw (76.center) to (78);
		\draw (78) to (69);
		\draw (77.center) to (74);
		\draw (74) to (70);
		\draw [bend left] (78) to (80.center);
		\draw [bend right] (78) to (83.center);
		\draw [bend right] (74) to (81.center);
		\draw [bend left] (74) to (82.center);
		\draw (76.center) to (84.center);
		\draw (84.center) to (85.center);
		\draw (85.center) to (77.center);
		\draw [bend right, looseness=1.25] (75) to (86.center);
		\draw [bend left] (75) to (71.center);
		\draw (75) to (85.center);
		\draw [bend right, looseness=1.25] (89) to (87.center);
		\draw [bend left] (89) to (88.center);
		\draw (84.center) to (89);
		\draw (67) to (89);
	\end{pgfonlayer}
\end{tikzpicture}
+
\begin{tikzpicture} 
	\begin{pgfonlayer}{nodelayer}
		\node [style={dummy}] (67) at (-2, 1.5) {$x$};
		\node [style={dummy}] (68) at (-2, -1.5) {$x$};
		\node [style={white_circle}] (69) at (1, 1.5) {$a$};
		\node [style={dummy}] (70) at (1, -1.5) {$x$};
		\node [style=none] (71) at (-1.25, -0.25) {};
		\node [style=none] (72) at (-1, -0.5) {\scriptsize 1};
		\node [style=none] (73) at (-1, 0.5) {\scriptsize 2};
		\node [style=tri] (74) at (0.5, -1) {};
		\node [style=tri] (75) at (-1.5, -1) {};
		\node [style=none] (76) at (0.25, 0.75) {};
		\node [style=none] (77) at (0.25, -0.75) {};
		\node [style=lam] (78) at (0.5, 1) {};
		\node [style=none] (79) at (0, 0.5) {\scriptsize 3};
		\node [style=none] (80) at (0.25, 0.25) {};
		\node [style=none] (81) at (0.25, -0.25) {};
		\node [style=none] (82) at (-0.25, -0.75) {};
		\node [style=none] (83) at (-0.25, 0.75) {};
		\node [style=none] (84) at (-1.25, 0.75) {};
		\node [style=none] (85) at (-1.25, -0.75) {};
		\node [style=none] (86) at (-0.75, -0.75) {};
		\node [style=none] (87) at (-1.25, 0.25) {};
		\node [style=none] (88) at (-0.75, 0.75) {};
		\node [style=tri] (89) at (-1.5, 1) {};
		\node [style=none] (90) at (0, -0.5) {};
		\node [style=none] (91) at (0, -0.5) {\scriptsize 4};
		\node [style=none] (92) at (0.25, 0.25) {};
	\end{pgfonlayer}
	\begin{pgfonlayer}{edgelayer}
		\draw (75) to (68);
		\draw (76.center) to (77.center);
		\draw (76.center) to (78);
		\draw (78) to (69);
		\draw (77.center) to (74);
		\draw (74) to (70);
		\draw [bend left] (78) to (80.center);
		\draw [bend right] (78) to (83.center);
		\draw [bend right] (74) to (81.center);
		\draw [bend left] (74) to (82.center);
		\draw (76.center) to (84.center);
		\draw (84.center) to (85.center);
		\draw (85.center) to (77.center);
		\draw [bend right, looseness=1.25] (75) to (86.center);
		\draw [bend left] (75) to (71.center);
		\draw (75) to (85.center);
		\draw [bend right, looseness=1.25] (89) to (87.center);
		\draw [bend left] (89) to (88.center);
		\draw (84.center) to (89);
		\draw (67) to (89);
	\end{pgfonlayer}
\end{tikzpicture}
+
\begin{tikzpicture} 
	\begin{pgfonlayer}{nodelayer}
		\node [style={dummy}] (67) at (-2, 1.5) {$x$};
		\node [style={dummy}] (68) at (-2, -1.5) {$x$};
		\node [style={dummy}] (69) at (1, 1.5) {$x$};
		\node [style={white_circle}] (70) at (1, -1.5) {$a$};
		\node [style=none] (71) at (-1.25, -0.25) {};
		\node [style=none] (72) at (-1, -0.5) {\scriptsize 1};
		\node [style=none] (73) at (-1, 0.5) {\scriptsize 2};
		\node [style=lam] (74) at (0.5, -1) {};
		\node [style=tri] (75) at (-1.5, -1) {};
		\node [style=none] (76) at (0.25, 0.75) {};
		\node [style=none] (77) at (0.25, -0.75) {};
		\node [style=tri] (78) at (0.5, 1) {};
		\node [style=none] (79) at (0, 0.5) {\scriptsize 3};
		\node [style=none] (80) at (0.25, 0.25) {};
		\node [style=none] (81) at (0.25, -0.25) {};
		\node [style=none] (82) at (-0.25, -0.75) {};
		\node [style=none] (83) at (-0.25, 0.75) {};
		\node [style=none] (84) at (-1.25, 0.75) {};
		\node [style=none] (85) at (-1.25, -0.75) {};
		\node [style=none] (86) at (-0.75, -0.75) {};
		\node [style=none] (87) at (-1.25, 0.25) {};
		\node [style=none] (88) at (-0.75, 0.75) {};
		\node [style=tri] (89) at (-1.5, 1) {};
		\node [style=none] (90) at (0, -0.5) {};
		\node [style=none] (91) at (0, -0.5) {\scriptsize 4};
		\node [style=none] (92) at (0.25, 0.25) {};
	\end{pgfonlayer}
	\begin{pgfonlayer}{edgelayer}
		\draw (75) to (68);
		\draw (76.center) to (77.center);
		\draw (76.center) to (78);
		\draw (78) to (69);
		\draw (77.center) to (74);
		\draw (74) to (70);
		\draw [bend left] (78) to (80.center);
		\draw [bend right] (78) to (83.center);
		\draw [bend right] (74) to (81.center);
		\draw [bend left] (74) to (82.center);
		\draw (76.center) to (84.center);
		\draw (84.center) to (85.center);
		\draw (85.center) to (77.center);
		\draw [bend right, looseness=1.25] (75) to (86.center);
		\draw [bend left] (75) to (71.center);
		\draw (75) to (85.center);
		\draw [bend right, looseness=1.25] (89) to (87.center);
		\draw [bend left] (89) to (88.center);
		\draw (84.center) to (89);
		\draw (67) to (89);
	\end{pgfonlayer}
\end{tikzpicture}
\right)
\right],
\label{eq:beta_tad_ctb_graph}
\end{align}
where the orange vertex corresponds the tadpole parameter $t_a$ of the Lagrangian \eqref{eq:Lag_general}. The analytic form of eq.~\eqref{eq:beta_tad_ctb_graph} looks like 
\begin{align}
	\beta_{a} & = 
		  t_{f} \gamma^\phi_{fb} 
		     + 
			     h_{aef} m^2_{gh} 
		     \left[
				     \bigcirc_{ef|gh}
				     +
				     \bigcirc_{gh|ef}
		     \right]
	\nonumber\\
		     & +
		     \left[
			     h_{aef} h_{ghi} h_{jkl}
		     	      \bigtriangleup_{ef|ghi|jkl}
			      + 
			      m^2_{ef} \lambda_{aghi} h_{jkl}
			     \left(
				     \bigtriangleup_{ef|ghi|jkl}
				     +
				     \bigtriangleup_{ef|jkl|ghi}
	     		\right)
		     \right]
		     \nonumber\\
		     & 
		     +
			     \lambda_{aefg} h_{hij} h_{klm} h_{nop}
		     \left[
		          \square_{efg|hij|klm|nop} 
		         +\square_{nop|efg|hij|klm} 
			 \right.
			 \nonumber\\
		     & 
		     \left.
		         +\square_{klm|nop|efg|hij} 
		         +\square_{hij|klm|nop|efg} 
		     \right].
		     \label{eq:beta_tad_ctb}
\end{align}
One more contraction with the dummy field $x_a$ gives the beta function of the vacuum energy:
\begin{align}
	\beta_{\Lambda} & =  
\left[
\begin{tikzpicture}
	\begin{pgfonlayer}{nodelayer}
		\node [style={dummy}] (0) at (-2.25, 1.5) {$x$};
		\node [style={dummy}] (1) at (-2.25, -1.5) {$x$};
		\node [style={dummy}] (2) at (2.25, 1.5) {$x$};
		\node [style={dummy}] (3) at (2.25, -1.5) {$x$};
		\node [style=none] (4) at (-0.75, 0) {};
		\node [style=none] (5) at (0.75, 0) {};
		\node [style=none] (6) at (-0.5, 0) {\scriptsize 1};
		\node [style=none] (7) at (0.5, 0) {\scriptsize 2};
		\node [style=mm] (18) at (1.25, 0) {};
		\node [style=mm] (21) at (-1.25, 0) {};
		\node [style=none] (57) at (-0.75, 0.25) {};
		\node [style=none] (58) at (-0.75, -0.25) {};
		\node [style=none] (59) at (0.75, -0.25) {};
		\node [style=none] (60) at (0.75, 0.25) {};
	\end{pgfonlayer}
	\begin{pgfonlayer}{edgelayer}
		\draw [bend left=90, looseness=1.50] (4.center) to (5.center);
		\draw [bend right=90, looseness=1.75] (4.center) to (5.center);
		\draw (2) to (18);
		\draw (18) to (3);
		\draw (0) to (21);
		\draw (21) to (1);
		\draw [bend left, looseness=0.75] (21) to (57.center);
		\draw [bend right] (21) to (58.center);
		\draw [bend right] (18) to (60.center);
		\draw [bend left] (18) to (59.center);
	\end{pgfonlayer}
\end{tikzpicture}
+
\begin{tikzpicture}
	\begin{pgfonlayer}{nodelayer}
		\node [style={dummy}] (22) at (-2, 1.5) {$x$};
		\node [style={dummy}] (23) at (-2, -1.5) {$x$};
		\node [style={dummy}] (24) at (1.75, 1.5) {$x$};
		\node [style={dummy}] (25) at (1.75, -1.5) {$x$};
		\node [style=none] (26) at (-0.5, 0) {};
		\node [style=none] (28) at (0, 0) {\scriptsize 1};
		\node [style=none] (29) at (0.75, 0.25) {\scriptsize 2};
		\node [style=tri] (30) at (1.25, -1) {};
		\node [style=mm] (31) at (-0.75, 0) {};
		\node [style=none] (32) at (1, 0.75) {};
		\node [style=none] (33) at (1, -0.75) {};
		\node [style=tri] (34) at (1.25, 1) {};
		\node [style=none] (36) at (0.75, -0.25) {\scriptsize 3};
		\node [style=none] (61) at (1, 0.25) {};
		\node [style=none] (62) at (1, -0.25) {};
		\node [style=none] (63) at (0.5, -0.5) {};
		\node [style=none] (64) at (0.5, 0.5) {};
		\node [style=none] (65) at (0, 0.25) {};
		\node [style=none] (66) at (0, -0.25) {};
	\end{pgfonlayer}
	\begin{pgfonlayer}{edgelayer}
		\draw (22) to (31);
		\draw (31) to (23);
		\draw (26.center) to (32.center);
		\draw (32.center) to (33.center);
		\draw (33.center) to (26.center);
		\draw (32.center) to (34);
		\draw (34) to (24);
		\draw (33.center) to (30);
		\draw (30) to (25);
		\draw [bend left, looseness=1.25] (31) to (65.center);
		\draw [bend right, looseness=1.25] (31) to (66.center);
		\draw [bend left] (34) to (61.center);
		\draw [bend right] (34) to (64.center);
		\draw [bend right] (30) to (62.center);
		\draw [bend left] (30) to (63.center);
	\end{pgfonlayer}
\end{tikzpicture}
+
\begin{tikzpicture} 
	\begin{pgfonlayer}{nodelayer}
		\node [style={dummy}] (67) at (-2, 1.5) {$x$};
		\node [style={dummy}] (68) at (-2, -1.5) {$x$};
		\node [style={dummy}] (69) at (1, 1.5) {$x$};
		\node [style={dummy}] (70) at (1, -1.5) {$x$};
		\node [style=none] (71) at (-1.25, -0.25) {};
		\node [style=none] (72) at (-1, -0.5) {\scriptsize 1};
		\node [style=none] (73) at (-1, 0.5) {\scriptsize 2};
		\node [style=tri] (74) at (0.5, -1) {};
		\node [style=tri] (75) at (-1.5, -1) {};
		\node [style=none] (76) at (0.25, 0.75) {};
		\node [style=none] (77) at (0.25, -0.75) {};
		\node [style=tri] (78) at (0.5, 1) {};
		\node [style=none] (79) at (0, 0.5) {\scriptsize 3};
		\node [style=none] (80) at (0.25, 0.25) {};
		\node [style=none] (81) at (0.25, -0.25) {};
		\node [style=none] (82) at (-0.25, -0.75) {};
		\node [style=none] (83) at (-0.25, 0.75) {};
		\node [style=none] (84) at (-1.25, 0.75) {};
		\node [style=none] (85) at (-1.25, -0.75) {};
		\node [style=none] (86) at (-0.75, -0.75) {};
		\node [style=none] (87) at (-1.25, 0.25) {};
		\node [style=none] (88) at (-0.75, 0.75) {};
		\node [style=tri] (89) at (-1.5, 1) {};
		\node [style=none] (90) at (0, -0.5) {};
		\node [style=none] (91) at (0, -0.5) {\scriptsize 4};
		\node [style=none] (92) at (0.25, 0.25) {};
	\end{pgfonlayer}
	\begin{pgfonlayer}{edgelayer}
		\draw (75) to (68);
		\draw (76.center) to (77.center);
		\draw (76.center) to (78);
		\draw (78) to (69);
		\draw (77.center) to (74);
		\draw (74) to (70);
		\draw [bend left] (78) to (80.center);
		\draw [bend right] (78) to (83.center);
		\draw [bend right] (74) to (81.center);
		\draw [bend left] (74) to (82.center);
		\draw (76.center) to (84.center);
		\draw (84.center) to (85.center);
		\draw (85.center) to (77.center);
		\draw [bend right, looseness=1.25] (75) to (86.center);
		\draw [bend left] (75) to (71.center);
		\draw (75) to (85.center);
		\draw [bend right, looseness=1.25] (89) to (87.center);
		\draw [bend left] (89) to (88.center);
		\draw (84.center) to (89);
		\draw (67) to (89);
	\end{pgfonlayer}
\end{tikzpicture}
\right]
\label{eq:beta_vac_ctb_graph}
\end{align}
corresponding to
\begin{align}
	\beta_{\Lambda} & = 
			     m^2_{ef} m^2_{gh} 
				     \bigcirc_{ef|gh}
			     + m^2_{ef} h_{ghi} h_{jkl}
		     	      \bigtriangleup_{ef|ghi|jkl}
			     + h_{efg} h_{hij} h_{klm} h_{nop}
		              \square_{efg|hij|klm|nop}.
\label{eq:beta_vac_ctb}
\end{align}
Loop expansion of the structures
\begin{align}
	\bigcirc_{ab|cd} & = \sum\limits_{l=1}^{6} \bigcirc^{(l)}_{ab|cd}, \\
			 \bigtriangleup_{ab|cde|fgh} & = \sum\limits_{l=1}^{6} \bigtriangleup^{(l)}_{ab|cde|fgh},\\
			 \square_{abc|def|ghi|jkl} & = \sum\limits_{l=1}^{6} \square^{(l)}_{abc|def|ghi|jkl}.
\end{align}
can be found in a supplementary PDF file.
\bibliographystyle{JHEP}
\bibliography{phi4gen6l}
\end{document}